\documentclass[usenatbib]{mn2e}
\usepackage{natbibmnfix, graphicx, times, amsmath, epsfig, amssymb}

\newcommand{\cmfast}{\textsc{\small 21CMFAST}}

\newcommand{\nf}{x_{\rm HI}}
\newcommand{\avenf}{\bar{x}_{\rm HI}}

\newcommand{\lya}{Ly$\alpha$}

\newcommand{\delT}{\delta T_b}

\newcommand{\Mmin}{M_{\rm min}}

\newcommand\lsim{\mathrel{\rlap{\lower4pt\hbox{\hskip1pt$\sim$}}
        \raise1pt\hbox{$<$}}}
\newcommand\gsim{\mathrel{\rlap{\lower4pt\hbox{\hskip1pt$\sim$}}
        \raise1pt\hbox{$>$}}}
\def\myputfigure#1#2#3#4#5%
{\vskip#5pt\makebox[0pt]{\hskip#2in
\includegraphics[width=#3\textwidth]{#1}}\vskip#4pt\hfill}

\newenvironment{packed_enum}{
\begin{enumerate}
  \setlength{\itemsep}{1pt}
  \setlength{\parskip}{0pt}
  \setlength{\parsep}{0pt}
}{\end{enumerate}}

\pdfoutput=1

\begin{document}

\title[Inhomogeneous recombinations during cosmic reionization]{Inhomogeneous recombinations during cosmic reionization}
\author[Sobacchi \& Mesinger]{Emanuele Sobacchi$^1$\thanks{email: emanuele.sobacchi@sns.it}, Andrei Mesinger$^1$  \\
$^1$Scuola Normale Superiore, Piazza dei Cavalieri 7, 56126 Pisa, Italy\\
}

\voffset-.6in

\maketitle

\begin{abstract}
By depleting the ionizing photon budget available to expand cosmic HII regions, recombining systems (or Lyman limit systems) can have a large impact during (and following) cosmic reionization.   Unfortunately, directly resolving such structures in large-scale reionization simulations is computationally impractical.  Instead, here we implement a sub-grid prescription for tracking inhomogeneous recombinations in the intergalactic medium.  Building on previous work parameterizing photo-heating feedback on star-formation, we present large-scale, semi-numeric reionization simulations which self-consistently track the local (sub-grid) evolution of {\it both} sources {\it and} sinks of ionizing photons.
Our simple, single-parameter model naturally results in both an extended reionization and a modest, slowly-evolving emissivity, consistent with observations.
Recombinations are instrumental in slowing the growth of large HII regions, and damping the rapid rise of the ionizing background in the late stages of (and following) reionization.  As a result, typical HII regions are smaller by factors of $\sim$2--3 throughout reionization.  
The large-scale ($k\lsim0.2$ Mpc$^{-1}$) ionization power spectrum is suppressed by factors of $\gsim2$--3 in the second half of reionization.
Therefore properly modeling recombinations is important in interpreting virtually all reionization observables, including upcoming interferometry with the redshifted 21cm line.  Consistent with previous works, we find the clumping factor of ionized gas to be $C_{\rm HII}\sim4$ at the end of reionization.
\end{abstract}


\section{Introduction}
\label{sec:intro}

Reionization is a story of sources and sinks of ionizing photons. Historically, the most famous protagonists of this tale have been the sources: the tiny ($\lsim$ percent level) fraction of baryons which are able to condense and form stars inside the first galaxies. A fraction of the ionizing photons emitted from these stars (and accreting black holes) manage to escape their host galaxies and begin ionizing the surrounding intergalactic medium (IGM). Driven by the birth and evolution of galaxies, these cosmic HII regions grow and overlap, eventually permeating all of space and completing the last major phase change of our Universe: hydrogen reionization.

On the other hand, the ``sinks'' of ionizing photons have received comparably less attention. An ionizing photon escaping a galaxy can have one of two fates: (i) it can ionize a neutral atom beyond the surrounding HII region, contributing to its growth; or (ii) while passing through the surrounding HII region it can encounter and reionize an atom which was previously ionized but subsequently recombined.  These later sinks (often dominated by so-called Lyman limit systems, LLSs, though they can also include less dense, more diffuse structures) could substantially delay reionization by depleting the photon budget available for expanding the HII regions (e.g. \citealt{MHR00, CSSF06, QLZD07, FOOD12, KG13}).  Furthermore, sinks can substantially alter reionization morphology, imposing a limit to the size of isolated HII regions (e.g. \citealt{FO05, MMS12, AA12}).  Understanding their impact is crucial in interpreting all observations of the epoch of reionization.

LLSs are better understood at lower redshifts ($z\lsim4$), post-reionization. Radiative transfer calculations \citep{QOF11, RPRS13}, together with observations of quasar $\text{Ly}\alpha$ spectra (e.g. \citealt{POW10, SC10}), confirm that the general properties of LLSs can be understood in terms of the local Jeans length \citep{Schaye01}, although the detailed properties are difficult to simulate as they (and especially even more dense systems) can occur very near galaxies.  Most IGM recombinations are sourced by gas with densities near the critical threshold required for self-shielding (e.g. \citealt{MHR00, FO05}).

Unfortunately, extending this simple picture of sinks to the epoch of reionization is not straightforward.  From a modeling standpoint, as always the main difficulty lies in the enormous dynamical range required. One needs to fully resolve the internal structure of the dominant population of self-shielded regions (with typical sizes of $\sim$ ten proper kpc; e.g. \citealt{Schaye01})\footnote{If the IGM is cold prior to reionization, the Jeans length could even be 1--2 orders of magnitude smaller (e.g. \citealt{ETA13}).} in cosmological simulations which sample reionization structure on scales of $\gtrsim$ hundreds of comoving Mpc (e.g. \citealt{FHZ04}).  Although single-box simulations of reionization have made significant progress lately (for example, see the recent review in \citealt{TG11}), tackling this full range of scales remains computationally out of reach for the foreseeable future.

Moreover, the properties of the recombining gas are sensitive to the local density and radiation fields, both of which can be very inhomogeneous and can evolve rapidly during reionization.  Conclusions regarding the role of sinks must therefore be robust to astrophysical uncertainties in these quantities.

Here we implement a sub-grid model of recombinations inside large-scale, semi-numerical simulations of reionization.  Each cell's recombination history depends on the local values of density, ionizing UV background (UVB), and reionization history.  Building on \citet{SM13b}, our simulations also include the effects of photo-heating on the star formation rate; therefore, we include reionization feedback on both the sources and the sinks.  We investigate the relative impact of inhomogeneous recombinations on the reionization history and morphology, as well as the evolution of the UVB, emissivity, mean free path to ionizing photons and the clumping factor (see below) of the ionized gas.

This paper is organized as follows. In Section \ref{sec:reion} we describe our large-scale reionization simulations.  In Section \ref{sec:subgrid} we present our sub-grid prescriptions to model UVB feedback on galaxies (\S \ref{sec:UVB_feedback}), and the small-scale structure of the IGM (\S \ref{sec:fluct}). In Section \ref{sec:results} we present our results, investigating the role of sinks and sources on IGM properties during reionization.  In Section \ref{sec:concl} we present our conclusions.  Throughout we assume a ﬂat $\Lambda$CDM cosmology with parameters ($\Omega_{\rm m}$, $\Omega_{\rm \Lambda}$, $\Omega_{\rm b}$, $h$, $\sigma_{\rm 8}$, $n$) = ($0.28$, $0.72$, $0.046$, $0.70$, $0.82$, $0.96$), as measured by the Wilkinson Microwave Anisotropy Probe (WMAP; \citealt{WMAP13}), and also consistent with recent results from the Planck satellite \citep{Planck13}. Unless stated otherwise, we quote all quantities in comoving units.

\section{Large-scale simulations of reionization}
\label{sec:reion}

To model cosmological reionization, we use a parallelized version of the publicly available semi-numerical simulation, \cmfast\footnote{http://homepage.sns.it/mesinger/Sim}. We generate the IGM density and source fields by: (i) creating a 3D Monte Carlo realization of the linear density field in a box with sides $L=300\text{ Mpc}$ and $N=1600^{3}$ grid cells; (ii) evolving the density field using the Zeldovich approximation \citep{Zeldovich70}, and smoothing onto a lower-resolution $N=400^3$ grid; (iii) using excursion-set theory \citep{PS74, BCEK91, LC93, ST99} on the evolved density field to compute the fraction of matter collapsed in halos bigger than a threshold $M_{\rm min}$ (see Section \ref{sec:UVB_feedback}), thus contributing to reionization (see \citealt{MF07, MFC11} for a more detailed description of the code).\footnote{We perform a rudimentary check of resolution convergence by redoing our main runs on a coarser, $N=200^3$ grid.  We recover all of the trends predicted by our fiducial resolution 
runs, in particular finding that all reionization histories are affected by less then $\Delta z \lsim 0.4$.}

The ionization field is computed by comparing the integrated number of ionizing photons to the number of baryons plus recombinations, in spherical regions of decreasing radius $R$ (i.e. following the excursion-set approach of \citealt{FHZ04}).
Specifically, a cell located at spatial position and redshift, ($\textbf{x}$, $z$), is flagged as ionized if:
\begin{equation}
\label{eq:ion_crit_coll}
\xi f_{\rm coll}(\textbf{x}, z, R, \bar{M}_{\rm min})\geq 1+\bar{n}_{\rm rec}(\textbf{x}, z, R)
\end{equation}
where $f_{\rm coll}\left(\textbf{x},z, R, \bar{M}_{\rm min}\right)$ is the fraction of collapsed matter inside a sphere of radius $R$ residing in halos larger than $\bar{M}_{\rm min}$, and $\xi$ is an ionizing efficiency, defined below.
 As we detail below, each cell keeps track of the local values of $M_{\rm min}($\textbf{x}$, $z$)$ and $n_{\rm rec}($\textbf{x}$, $z$)$, computed according to the cell's density and ionization history.  {\it The latter is the main improvement of this work}. When computing the ionization criterion in eq. (\ref{eq:ion_crit_coll}), $\bar{M}_{\rm min}$ and $\bar{n}_{\rm rec}$ are also averaged over scale $R$.\footnote{As discussed in \citet{SM13b}, formally we should be averaging over the local values of $f_{\rm coll}$, instead of $M_{\rm min}$, i.e. $\langle f_{\rm coll}(M_{\rm min}) \rangle_R \neq f_{\rm coll}(\langle M_{\rm min} \rangle_R)$. However, only computing $f_{\rm coll}$ locally in each cell could yield spurious, resolution-dependent results, since we are using the evolved (instead of the linear) density field: an application of the conditional mass function which has only been tested empirically within the context of the excursion-set approach to reionization \citep{ZMQT11, MFC11}.  In \citet{SM13b} 
we show that under maximally pessimistic assumptions, this approach results in a mis-estimate of the effective collapse fraction of at most 10--20\%, comparable to the uncertainty in the high-redshift mass function itself.}

Starting from the box size, the smoothing scale is decreased, and the criterion in eq. (\ref{eq:ion_crit_coll}) is re-evaluated.   It is important to note that most previous excursion-set approaches add a maximum starting scale, $R_{\rm max}$, generally corresponding to a chosen value for the mean free path to ionizing photons through the ionized IGM, $\lambda_{\rm mfp, HII} = R_{\rm max}$.  This value is usually treated as homogeneous and redshift independent.  In this work we remove this free parameter, as our procedure explicitly computes the local mean free path. 
At the cell size, $R_{\rm cell}$, the partial ionizations from sub-grid sources are evaluated, and the cell's ionized fraction is set to $\xi f_{\rm coll}({\bf x}, z, R_{\rm cell}, M_{\rm min})/({1+n_{\rm rec}})$ \citep{MFC11}. Neglecting recombinations, this algorithm results in ionization fields which are in good agreement with cosmological radiative transfer algorithms on $\gsim$ Mpc scales \citep{ZMQT11}.

The ionizing efficiency from eq. (\ref{eq:ion_crit_coll}) can be written out as:
\begin{equation}
\label{eq:zeta}
\xi = 30 \bigg(\frac{N_\gamma}{4000}\bigg) \bigg(\frac{f_{\rm esc}}{0.15}\bigg) \bigg(\frac{f_\ast}{0.05}\bigg) \bigg(\frac{f_{\rm b}}{1}\bigg)~ .
\end{equation}
where $N_\gamma$ is the number of ionizing photons per stellar baryon, $f_{\rm esc}$ is the fraction of UV ionizing photons that escape into the IGM, $f_\ast$ is the fraction of galactic gas in stars, and $f_{\rm b}$ is the fraction of baryons inside the galaxy with respect to the cosmic mean $\Omega_{\rm b}/\Omega_{\rm m}$ (note that some works include the total number of homogeneous recombinations inside the definition of $\xi$).
Although our models depend only on the product in eq. (\ref{eq:zeta}), we show on the RHS some reasonable values for the component terms.  $N_\gamma \approx 4000$ is expected for PopII stars (e.g. \citealt{BL05_WF}). On the other hand, the parameters $f_{\rm esc}$,  $f_\ast$ and $f_{\rm b}$ are extremely uncertain in high-redshift galaxies (e.g. \citealt{GKC08, WC09, FL13}), though our fiducial choices are in agreement with high-redshift galaxy luminosity functions (e.g. \citealt{Robertson13})\footnote{There are likely $\sim$two orders of magnitude separating the halo masses corresponding to the atomic cooling threshold and the current sensitivity limits of high-redshift ($z\sim 6-10$) Lyman break galaxy (LBGs) surveys (e.g. \citealt{ML11, SFD11, KF12}).
Thus the observed LBG candidates likely do not correspond to the dwarf galaxies which dominate reionization (e.g. \citealt{CFG08, AFT12}).
Furthermore there are several uncertainties when mapping the observed UV luminosity to the ionizing UV luminosity (e.g. \citealt{KF12}). Hence, we do not use the observed candidate LBGs to directly constrain our reionizing source models (c.f. \citealt{KG13_LF}).  Instead, we draw general conclusions focusing on observations of the IGM, which are in fact sensitive to the bulk of the reionizing sources.}.

\subsection{Incorporating an Inhomogeneous UVB}
\label{sec:self_cons}

In order to determine the ionization state of the IGM, it is necessary to know $M_{\rm min}$, $\lambda_{\rm mfp}$ and $n_{\rm rec}$ (eq. \ref{eq:M_crit}, \ref{eq:mfp} and \ref{eq:dn_rec} respectively), which all depend on the photoionization rate in the local HII region. The mean emissivity (number of ionizing photons emitted into the IGM per unit time per baryon) can be written as:
\begin{equation}
\label{eq:emissivity}
\epsilon \approx f_{\rm b} f_\ast f_{\rm esc} N_{\gamma} \frac{df_{\rm coll}}{dt} \approx \xi  \frac{df_{\rm coll}}{dt} ~ ,
\end{equation}
The photoionization rate is proportional to the local ionizing mean free path, $\lambda_{\rm mfp}$ (set by {\it either} recombinations {\it or} reionization morphology), multiplied by the emissivity. Assuming that the emissivity is spectrally distributed as $\nu^{-\alpha}$ (in this paper we use $\alpha=5$, corresponding to a stellar-driven UV spectrum; e.g. \citealt{TW96}), we can write the average photoionization rate {\it in a given HII region} as
\begin{equation}
\label{eq:self_cons_J}
\Gamma_{\rm HII}(\textbf{x}, z) \approx \left(1+z\right)^2 \lambda_{\rm mfp} \sigma_{\rm H}\frac{\alpha}{\alpha+\beta}\bar{n}_{\rm b}\xi\frac{df_{\rm coll}}{dt} ~ .
\end{equation}
where $\bar{n}_{\rm b}$ is the mean baryon number density inside the HII region, and we have assumed a photoionization cross-section $\sigma\left(\nu\right)=\sigma_{\rm H}\left(\nu/\nu_{\rm H}\right)^{-\beta}$ with $\sigma_{\rm H}=6.3\times 10^{-18}\text{ cm$^2$}$ and $\beta =2.75$.

Due to the clustering of dark matter halos, the relevant intensity at galaxy locations will be higher than the average UVB intensity inside the local HII region.  Therefore when calculating $M_{\rm crit}$ in eq. (\ref{eq:M_crit}), we use $\Gamma_{\rm halo, HII} =f_{\rm bias}\times \Gamma_{\rm HII}$, with $f_{\rm bias}=2$, consistent with results from \citet{MD08}.

\section{Implementing sub-grid physics}
\label{sec:subgrid}

Unlike previous studies, our approach self-consistently follows the local star-formation and recombination rates based on each cell's density, ionization history, and local UVB.  The recombination rate is computed using the full distribution of HI in the IGM.
Before we describe the details, we present the general steps of our algorithm.  Starting from a high-redshift (here taken to be $z_{\rm init}=15$) snapshot of the large-scale density field:
\begin{packed_enum}
\item Using the excursion-set procedure and criteria described above, we identify local HII regions around each cell, computing the associated UVB with $\lambda_{\rm mfp} = R$ (eq. \ref{eq:self_cons_J}).

\item If a cell is newly ionized, its ionization redshift and UVB intensity are recorded in order to account for photo-heating feedback on star formation, according to eq. (\ref{eq:M_crit}) \citep{SM13a}.

\item Treating each cell as a large-scale background (e.g. \citealt{CK89}), we model its density sub-structure with an empirical density distribution, calibrated to numerical simulations (eq. \ref{MHR_pdf}; \citealt{MHR00}).

\item Using the local photoionization rate, we compute the neutral fraction at each overdensity, assuming ionization equilibrium and a self-shielding prescription \citep{Schaye01, RPRS13}.

\item Integrating over these sub-structure distributions, we compute the recombination rate internal to the cell.  These recombinations are then added to the time-integrated total number of recombinations per baryon, $n_{\rm rec}(\textbf{x}, z)$ (eq. \ref{eq:dn_rec}-\ref{eq:n_rec}).

\item We move on to the next redshift snapshot of the density field, repeating the above steps with the updated quantities.
\end{packed_enum}

\subsection{UVB Feedback on Ionizing Sources}
\label{sec:UVB_feedback}

We implement UVB feedback as in \citet{SM13b}.  In this approach, the minimum mass of halos contributing to reionization is expressed in terms of two star-formation criteria: $M_{\rm min}=\max\left(M_{\rm cool}, M_{\rm crit}\right)$. $M_{\rm cool}$ corresponds to the virial temperature $T_{\rm vir}\approx10^{4}\text{ K}$ below which gas cannot cool through atomic hydrogen (molecular hydrogen cooling is expected to be strongly suppressed during the very early stages of reionization; e.g. \citealt{HRL97}). 

In reionized regions, the gas reservoir to form stars can be depleted from the additional heating from the UVB (lowering $f_b$ from eq. \ref{eq:zeta}; e.g. \citealt{SGB94, MR94, HG97}).  Halos with masses above $M_{\rm crit}$ retain enough gas (defined so that $f_b \geq 1/2$) to continue efficiently forming stars inside HII regions.  Using a functional form motivated by linear theory, in \citet{SM13a}, we obtained the following empirical formula:
\begin{equation}
\label{eq:M_crit}
M_{\rm crit}(\textbf{x}, z) = M_{0} \left(\frac{\Gamma_{\rm halo,HII}}{10^{-12}\text{ s$^{-1}$}}\right)^a \left(\frac{1+z}{10}\right)^{b}\left[1-\left(\frac{1+z}{1+z_{\rm IN}}\right)^{c}\right]^{d}
\end{equation}
where  $z$ is the collapse (i.e. current) redshift, $z_{\rm IN}$ is the redshift when the halo is first exposed to the UVB, and $\left(M_{0}, a, b, c, d\right) = \left(3.0\times 10^{9}M_{\odot}, 0.17, -2.1, 2.0, 2.5\right)$ are fitted to suites of 1D collapse simulations exploring a wide parameter space.  Eq. (\ref{eq:M_crit}) allows us to self-consistently keep track of the local value of $M_{\rm min}$ in each cell.

\subsection{Small Scale Structure of the IGM}
\label{sec:fluct}

We assume that the density sub-structure of each large-scale simulation cell is described according to the empirical formula, developed and calibrated to numerical simulations by \citet{MHR00}.   In this ``MHR'' model, the volume-weighted gas density distribution (i.e. the fraction of volume in which the gas is at overdensity $\Delta\equiv n_{\rm b}/\bar{n}_{\rm b}$) can be written as\footnote{We caution that the MHR density distribution is not calibrated to simulations at very high redshifts, nor at very high densities (which were not well-resolved by their simulation). This is approximately dealt with using an overall normalization, $f_{\rm s}$ (see eq. \ref{eq:CDDF2}).}:
\begin{equation}
\label{MHR_pdf}
P_{\rm V}\left(\Delta,z\right)=A\exp\left[-\frac{\left(\Delta^{-2/3}-C_{\rm 0}\right)^{2}}{2\left(2\delta_{\rm 0}/3\right)^{2}}\right]\Delta^{-\beta}
\end{equation}
where the fitted parameter $\delta_{\rm 0}=7.61/(1+z_{\rm eff})$ crudely scales as the Jeans length in the ionized IGM, which we evaluate at an effective redshift $(1+z_{\rm eff})\equiv \left(1+z\right)\Delta_{\rm cell}^{1/3}$; this is motivated by the self-similarity of the Einstein-de Sitter Universe, where each large-scale patch can be treated as a background Universe.\footnote{We are implicitly assuming that the density distribution of the gas responds instantaneously to photo-heating. Although this is not exactly the case \citep{PSS09}, this assumption is relatively safe since the response time-scale is shorter than the typical extension of inhomogeneous reionization.
Most importantly, it is likely the IGM is pre-heated by X-rays (e.g. \citealt{Oh01, RO04, MFS13}), dramatically reducing the time-scale for response. Nevertheless, the photon-consumption during gas relaxation remains an uncertainty.} 
The parameter $\beta$ is tabulated at $z<6$, while we assume $\beta =2.5$ at higher redshifts (corresponding to an isothermal sphere profile for high-density absorbers). $A$ and $C_{\rm 0}$ ensure volume and mass normalization of the distribution as appropriate for each cell's mean over-density, $\Delta_{\rm cell}$.

Assuming photoionization equilibrium, we calculate the neutral fraction at a given density:
\begin{equation}
\label{eq:nf}
\nf \Gamma_{\rm local} = \chi_{\rm HeII} ~ n_{\rm H} ~ (1-\nf)^2 ~  \alpha_{\rm B} ~ ,
\end{equation}
where $n_H = \Delta \bar{n}_{\rm H}$ is the hydrogen number density, $\alpha_{\rm B}=2.6\times 10^{-13}\text{ cm$^{3}$ s$^{-1}$}$ is the case B recombination coefficient for gas at $T\simeq 10^{4}\text{ K}$, and $\chi_{\rm HeII} = 1.08$ accounts for singly-ionized helium.  We take into account the self-shielding of the gas through a density-dependent photoionization rate, obtained by an empirical fit to radiative transfer simulations \citep{RPRS13}:
\begin{align}
\label{eq:UVB}
\frac{\Gamma_{\rm local}}{\Gamma_{\rm HII}} & =0.98\times\left[1+\left(\frac{\Delta}{\Delta_{\rm ss}}\right)^{1.64}\right]^{-2.28}+ \nonumber \\
& +0.02\times\left[1+\frac{\Delta}{\Delta_{\rm ss}}\right]^{-0.84}
\end{align}
where $\Delta_{\rm ss}$ is the overdensity above which the gas is self-shielded \citep{Schaye01}.  Using a spectrally-averaged ionization cross-section corresponding to our UVB power index of $\alpha=5$, we have:
\begin{equation}
\label{eq:D_ss}
\Delta_{\rm ss}=27\times\left(\frac{T}{10^{4}\text{ K}}\right)^{0.17}\left(\frac{1+z}{10}\right)^{-3}\left(\frac{\Gamma_{\rm HII}}{10^{-12}\text{ s$^{-1}$}}\right)^{2/3} ~.
\end{equation}

\begin{figure}
\vspace{+0\baselineskip}
{
\includegraphics[width=0.45\textwidth]{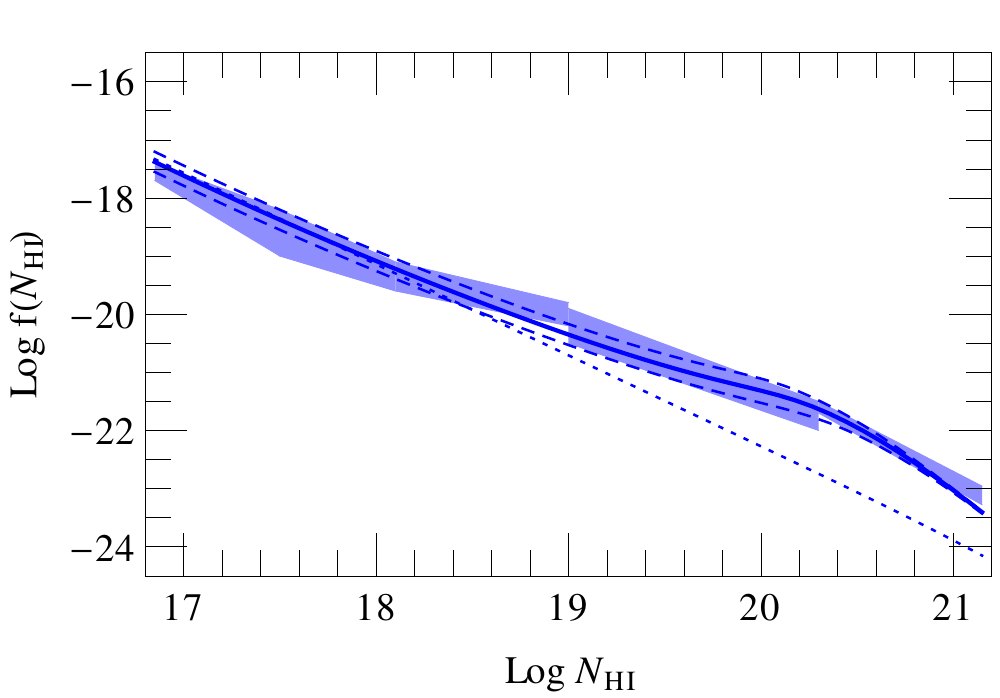}
}
\caption{CDDF of the gas calculated with different fiducial photoionization rates (in units of $10^{-12}\text{ s$^{-1}$}$). The shaded region corresponds to observational constraints at a mean redshift of $3.7$ (\citealt{OPB07, PW09, PWO09, POW10}; with the compilation from \citealt{QOF11}). Curves correspond to, \textit{solid}: $\Gamma_{\rm HII}=0.5$; \textit{dashed}: $\Gamma_{\rm HII}=0.25$ and $\Gamma_{\rm HII}=1$ (upper and lower curves respectively). The \textit{dotted} curve corresponds to the optically thin approximation with $\Gamma_{\rm HII}=0.5$.
\label{fig:CDDF}
}
\vspace{-1\baselineskip}
\end{figure}

To compare our model against observations we calculate the corresponding column density distribution function (CDDF) of the gas. This is usually expressed as the number of absorption lines per unit absorption distance $X$ (defined in eq. \ref{eq:CDDF}) and column density $N_{\rm HI}$. The relation of the CDDF to the observed number density of absorption lines per unit redshift $d^{2}n/dN_{\rm HI}dz$ depends on the assumed cosmology:
\begin{equation}
\label{eq:CDDF}
f\left(N_{\rm HI}, z\right)\equiv\frac{d^{2}n}{dN_{\rm HI} dX}\equiv\frac{d^{2}n}{dN_{\rm HI} dz}\frac{H\left(z\right)}{H_{\rm 0}}\frac{1}{\left(1+z\right)^{2}} ~ .
\end{equation}
We assume that the gas is in dynamical equilibrium and calculate $N_{\rm HI}$ as \citep{Schaye01}:
\begin{equation}
N_{\rm HI}=1.6\times 10^{21}\text{ cm$^{-2}$} n_{\rm H}^{1/2}\left(\frac{T}{10^{4}\text{ K}}\right)^{1/2}x_{\rm HI}
\end{equation}
Assuming uniform density absorbers, the CDDF is given by (\citealt{FO05}, Appendix A):
\begin{equation}
\label{eq:CDDF2}
f\left(N_{\rm HI},z\right)=f_{\rm s}\Omega_{\rm b}\frac{d\Delta}{dN_{\rm HI}}\Delta P_{\rm V}\left(\Delta,z\right)\frac{3H_{\rm 0}c\left(1-Y\right)}{8\pi Gm_{\rm H}}x_{\rm HI}N_{\rm HI}^{-1}
\end{equation}
Since the MHR distribution overestimates the amount of gas at densities $\Delta\gtrsim\Delta_{\rm ss}$ (e.g. \citealt{PSS09, BB09}), we re-scale eq. (\ref{eq:CDDF2}) by $f_{\rm s}\sim 0.3$ (given the uncertainty in the measured CDDF, $f_{\rm s}$ can not be constrained strictly; see also \citealt{FO05}, Appendix A).

In Figure \ref{fig:CDDF} we compare the resulting CDDF with observational constraints at a mean redshift of $3.7$ \citep{OPB07, PW09, PWO09, POW10}. We show the CDDF calculated assuming different fiducial photoionization rates (in units of $10^{-12}\text{ s$^{-1}$}$): $\Gamma_{\rm HII}=0.5$ (\textit{solid}), $\Gamma_{\rm HII}=0.25$ and $\Gamma_{\rm HII}=1$ (\textit{dashed}). The CDDF agrees well with observations, including the damped $\text{Ly}\alpha$ system (DLA) dip at $N_{\rm HI}\gtrsim 2\times 10^{20}\text{ cm$^{-2}$}$.
For comparison we also show the result of an optically thin approximation with $\Gamma_{\rm HII}=0.5$ (\textit{dotted}).  As expected, the optically thin approximation underestimates the CDDF at large column densities where the self-shielding of the gas becomes important.

\subsubsection{Recombination Rate}
\label{sec:rec}

\begin{figure}
\vspace{+0\baselineskip}
{
\includegraphics[width=0.45\textwidth]{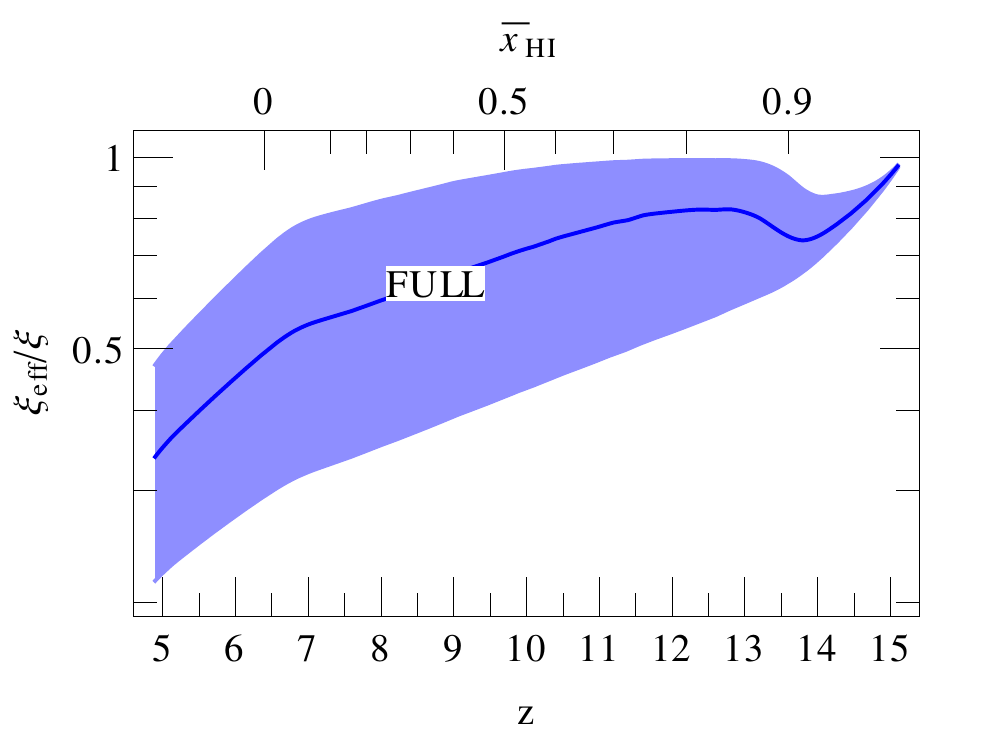}
}
\caption{Mean effective ionizing efficiency $\xi_{\rm eff}/\xi\equiv 1/\left(1+\bar{n}_{\rm rec}\right)$ in HII regions for the fiducial run, \textbf{FULL}. The shaded region corresponds to the 1-$\sigma$ spread among HII regions.
\label{fig:rec}
}
\vspace{-1\baselineskip}
\end{figure}

 Taking self-shielding into account, and integrating over the entire density distribution, the recombination rate per baryon in an ionized cell is:
\begin{equation}
\label{eq:dn_rec}
\frac{d n_{\rm rec}}{dt}(\textbf{x}, z)=\int_0^{+\infty}P_{\rm V}\left(\Delta, z\right)\Delta\bar{n}_{\rm H}\alpha_{\rm B}\left[1-x_{\rm HI}\left(\Delta\right)\right]^{2}d\Delta ~.
\end{equation}
Then the total, time-integrated number of recombinations per baryon, averaged over the smoothing scale (HII region), can be written as:
\begin{equation}
\label{eq:n_rec}
\bar{n}_{\rm rec}(\textbf{x}, z)=\left\langle\int_{z_{\rm IN}}^{z}\frac{dn_{\rm rec}}{dt}\frac{dt}{dz}dz\right\rangle_{\rm HII}
\end{equation}
\noindent Equation (\ref{eq:n_rec}) and its use in the standard reionization criterion, eq. (\ref{eq:ion_crit_coll}), represent the main modeling improvement of this work.

Recombinations consume UV photons, making the ionization criterion (eq. \ref{eq:ion_crit_coll}) more restrictive and slowing the propagation of HII fronts.  One way to think of this is in terms of a declining, ``effective'' ionizing efficiency, $\xi_{\rm eff}/\xi\equiv 1/\left(1+\bar{n}_{rec}\right)$.  We show the global impact of recombinations in terms of $\xi_{\rm eff}/\xi$ (averaged just over HII regions\footnote{We note that averaging just over HII regions results in a slight dip of $\left(1+\bar{n}_{rec}\right)^{-1}$ at $\avenf\sim0.9$.  This is likely caused by the fact that the first HII regions form around the highest density peaks, with the largest recombination rates.  Subsequent HII regions are less biased, flattening-out the ensemble average of $\left(1+\bar{n}_{rec}\right)^{-1}$.}) in Fig. \ref{fig:rec} for our fiducial model (described below). We see that recombinations have an important impact in the late stages of (and following) reionization, when the increasing UVB drives the 
ionization fronts into dense, rapidly-recombining systems.  For example, by the end of reionization ($z\approx6.5$ in Fig. \ref{fig:rec}), half of the emitted photons are consumed to balance recombinations.  As we shall see below, this drop in the effective ionizing efficiency cannot be balanced by a corresponding rapid rise in the star-formation rate $\propto df_{\rm coll}(>M_{\rm min})/dt$ (for reasonable models), resulting in a ``photon-starved'' or ``recombination-limited'' end to reionization.

\subsubsection{Mean Free Path of Ionizing Photons}
\label{sec:mfp}

\begin{figure}
\vspace{+0\baselineskip}
{
\includegraphics[width=0.45\textwidth]{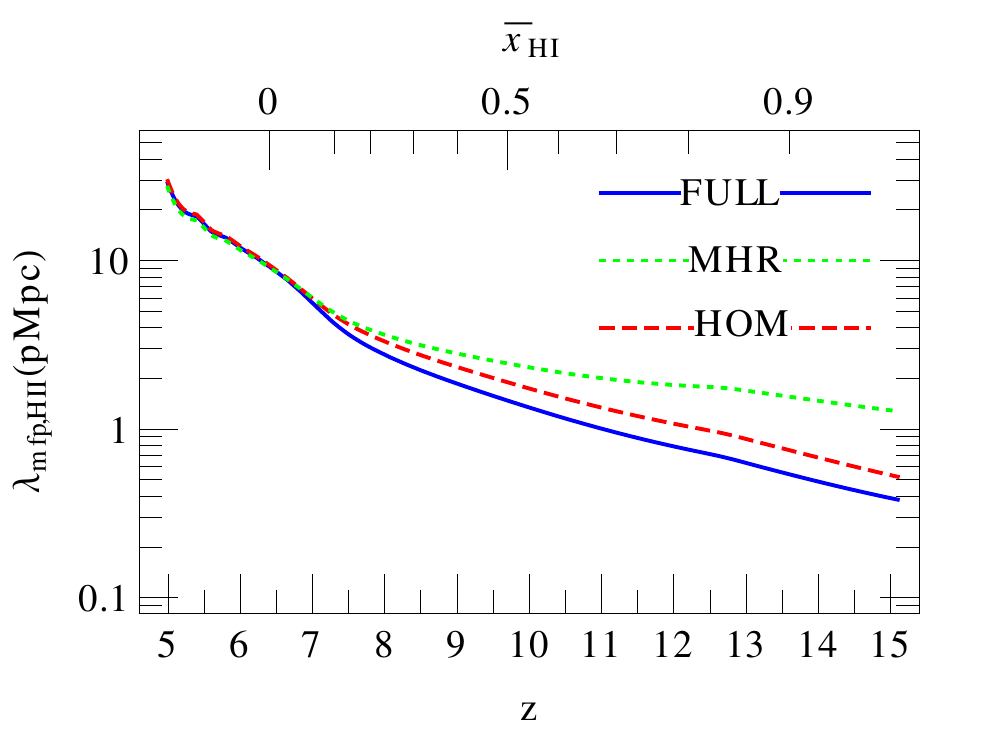}
}
\caption{Spatially-averaged mean free path (in proper Mpc) calculated using our complete model (eq. \ref{eq:mfp}; {\it blue solid curve}), ignoring the large-scale bias of HII regions ({\it red dashed curve}), and using the approximation of \citet{MHR00} (eq. \ref{eq:MHRmfp}; {\it green dotted curve}).  All models assume the same inhomogeneous UVB and reionization morphology, corresponding to our {\bf FULL} model.
\label{fig:mfp_comp}
}
\vspace{-1\baselineskip}
\end{figure}

\begin{figure*}
\vspace{+0\baselineskip}
{
\includegraphics[width=\textwidth,height=2.6cm]{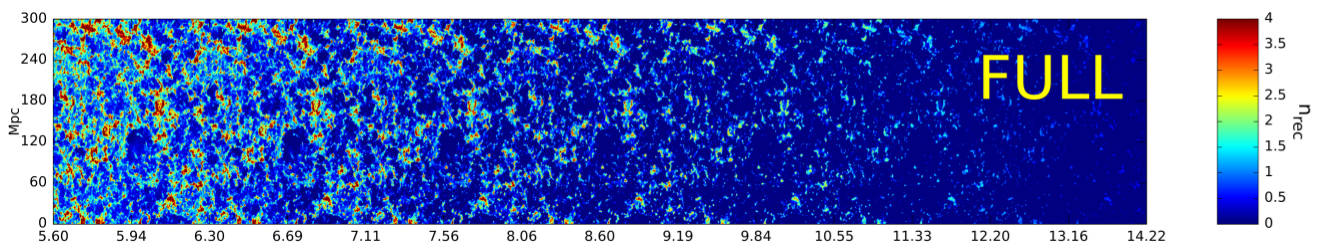}
}
\caption{Light-cone slice (0.75 Mpc thick) of the cumulative number of recombinations in the \text{FULL} model.
\label{fig:n_rec_box}
}
\vspace{-1\baselineskip}
\end{figure*}

\begin{figure*}
\vspace{+0\baselineskip}
{
\includegraphics[width=\textwidth,height=2.6cm]{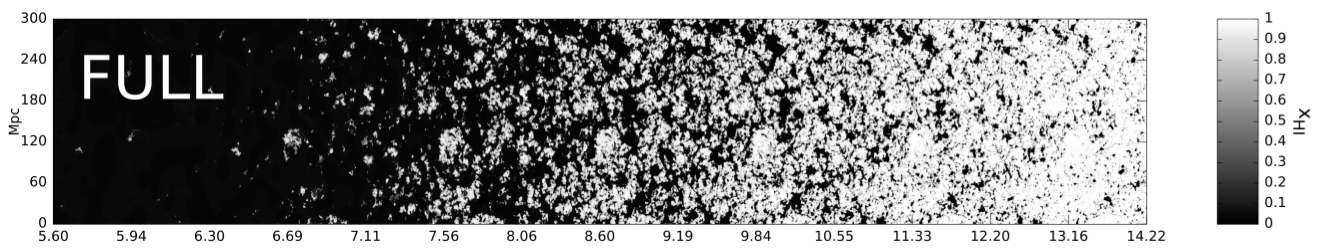}
\includegraphics[width=\textwidth,height=2.6cm]{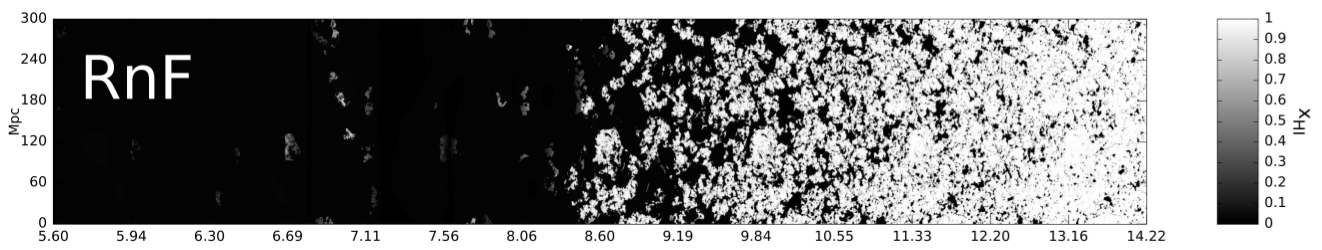}
\includegraphics[width=\textwidth,height=2.6cm]{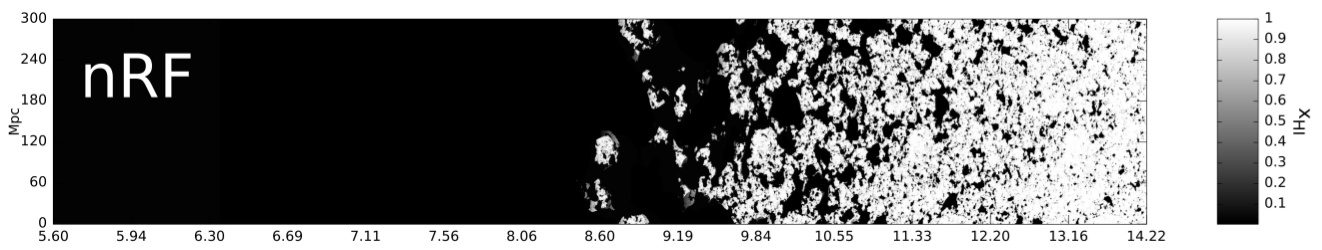}
\includegraphics[width=\textwidth,height=2.6cm]{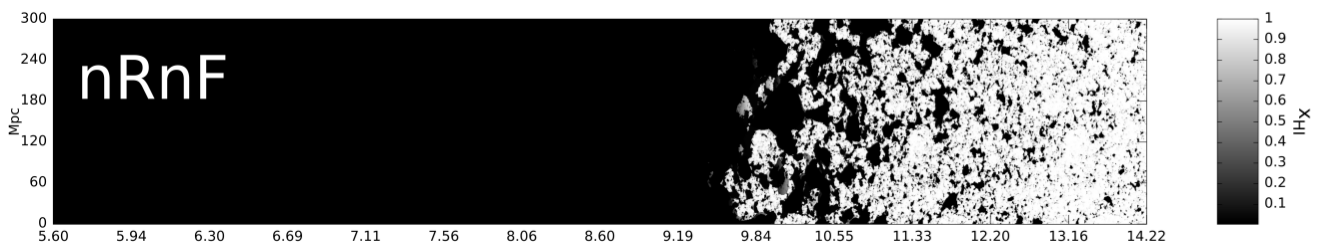}
}
\caption{Light-cone slices through our ionization boxes, with HII regions in black and HI regions in white. From top to bottom we show the runs: \textbf{FULL}, \textbf{RnF}, \textbf{nRF}, \textbf{nRnF}.  All slices are 0.75 Mpc thick.
\label{fig:light_box}
}
\vspace{-1\baselineskip}
\end{figure*}


Our formalism also allows us to compute the local mean free path through the {\it ionized} IGM, $\lambda_{\rm mfp, HII}$, \footnote{As mentioned above, the local mean free path, $\lambda_{\rm mfp}$ (e.g. eq. \ref{eq:self_cons_J}), can be set by either reionization morphology or by LLSs.  Generally, the former is important during the early stages of reionization when the HII regions are growing.  The later (i.e. $\lambda_{\rm mfp, HII}$) is important in the late stages of (and following) reionization, when HII regions locally become recombination limited, and most ionizing photons get absorbed by LLSs \citep{FO05, FM09, AA12}.} and compare with observational estimates post-reionization.  High-redshift predictions of $\lambda_{\rm mfp, HII}$ are also very useful in checking the empirical extrapolations used extensively in analytic models of reionization (e.g. \citealt{CF06}).

Using the CDDF from eq. (\ref{eq:CDDF2}), we can write \citep{FO05}:
\begin{equation}
\label{eq:mfp}
\lambda_{\rm mfp, HII} = \frac{c}{H_{\rm 0}\left(1+z\right)^{3}\int_{0}^{+\infty}f\left(N_{\rm HI}, z\right)\left(1-\frac{\Gamma_{\rm local}}{\Gamma_{\rm HII}}\right)dN_{\rm HI}}
\end{equation}
Assuming ionization equilibrium, $\lambda_{\rm mfp, HII}$ is set by the {\it instantaneous} recombination rate, and should be included in the time-integral of eq. (\ref{eq:ion_crit_coll}); for numerical robustness however we also require that $\lambda_{\rm mfp, HII}>R$ for a region to be ionized.  This added check (although we confirm that it has a negligible impact on our fiducial reionization history and morphology), helps ensure that rapidly-evolving recombination rates are not ``smoothed-over'' between two redshift snapshots.

Eq. (\ref{eq:mfp}) includes a weighted integral over the entire CDDF of the ionized IGM.  A simpler, two-phase model of the ionization state was proposed by \citet{MHR00}, where the neutral fraction of the local density patch effectively transitions from zero  at $\Delta < \Delta_{\rm ss}$ to unity at $\Delta > \Delta_{\rm ss}$.  Then the fraction of the IGM occupied by absorbers is $Q_{\rm ss} = \int_{\Delta_{\rm ss}}^\infty P(\Delta, z)d\Delta$. If absorbers are further assumed to correspond to discrete systems with a constant shape at a fixed $\Delta_{\rm ss}$, their radius is proportional to $Q_{\rm ss}^{1/3}$ and their mean separation to $Q_{\rm ss}^{-2/3}$.  In this simple picture, the mean free path can be written as:
\begin{equation}
\label{eq:MHRmfp}
\lambda_{\rm mfp, HII}=\lambda_{\rm 0}Q_{\rm ss}\left(\Delta_{\rm ss}\right)^{-2/3} ~ ,
\end{equation}
where $\lambda_{\rm 0}H=60\text{ km s$^{-1}$}$ is the chosen normalization, which implicitly also allows one to account for the physically non-negligible cumulative opacity of $\Delta < \Delta_{\rm ss}$ systems (\citealt{FO05}; Appendix A).

In Figure \ref{fig:mfp_comp} we compare the values of $\lambda_{\rm mfp, HII}$ calculated using the full CDDF (eq. \ref{eq:mfp}; blue solid curve) and the \citet{MHR00} approximation (eq. \ref{eq:MHRmfp}; green dotted curve).  Both models assume the same UVB and reionization morphology, corresponding to the {\bf FULL} model, described below.  At moderate redshifts ($z \lsim 7$), the results are surprisingly similar.  Some similarity between the two approaches is expected, given that the $\lambda_0$ normalization in eq. (\ref{eq:MHRmfp}) was chosen to match $z\sim3$ observations.  However, at $z \gsim 7$ the \citet{MHR00} approximation begins overestimating the mean free path, by up to a factor of $\sim 2$ at the highest redshifts we consider.  With the increasing mean density at higher redshifts, more common systems (lower $\Delta_{\rm ss}$) dominate the mean free path; hence the approximations of constant-shape, discretised absorbers inherent in eq. (\ref{eq:MHRmfp}) become increasingly inaccurate.

With the red dashed curve in Figure \ref{fig:mfp_comp}, we also include a calculation ignoring the modulation of the sub-grid density distribution by the large-scale density field when computing eq. (\ref{eq:mfp}); i.e. taking $z_{\rm eff}=z$ when computing $P(\Delta, z)$. In setting the spatially-averaged mean free path, the large-scale density inhomogeneities are only important at high redshifts when the HII regions are strongly biased (see Fig. \ref{fig:D_evo}).
Neglecting this density bias overestimates the mean free path by $\sim 20$\% at $z\gsim 8$.

\section{Results}
\label{sec:results}

We now investigate the relative impact of inhomogeneous recombinations on the reionization history, morphology, and other IGM properties.  Our fiducial model assumes $\xi=30$ and an atomic cooling threshold for star-forming galaxies ($M_{\rm cool}$ corresponding to $T_{\rm vir}\gsim 10^{4}\text{ K}$). This choice of $\xi$ is simply motivated by having reionization histories match the mean value of $\tau_{\rm e}$ as observed by WMAP.  This integral constraint still allows for a wide range of models; therefore the fiducial choice of $\xi$ should be treated merely as a rough ``guess''.  Keeping these parameters fixed, the runs we compare are:
\begin{itemize}
\item \textbf{FULL}: self-consistent calculation of $M_{\rm crit}$ and $n_{\rm rec}$, as described above.
\item \textbf{No Recombinations, Feedback (nRF)}: like \textbf{FULL}, but neglecting recombinations, i.e. $n_{\rm rec}=0$.
\item \textbf{Recombinations, No Feedback (RnF)}: like \textbf{FULL}, but neglecting UVB feedback on sources, i.e. $\Mmin = M_{\rm cool}$.
\item \textbf{No Recombinations, No Feedback (nRnF)}: like \textbf{FULL}, but neglecting both recombinations and UVB feedback on sources.
\end{itemize}



Obviously, these models are not meant to be exhaustive, spanning a wide range of astrophysical parameter space.  Rather they serve to further physical intuition about the impact and relative importance of inhomogeneous feedback from sources and sinks.  Eventually, more physical models can be constructed, including a halo mass and redshift dependence of source luminosities which is more complex than our simple $M_{\rm halo} > M_{\rm crit}({\bf x},z)$ threshold.

Before moving onto more quantitative analysis, in Fig. \ref{fig:n_rec_box} and Fig. \ref{fig:light_box}, we show light-cone slices through the recombination and ionization fields, respectively.  It is evident that both fields are quite inhomogeneous.  Large-scale overdensities are first to ionize, and these regions host the most recombinations.  In general, the large-scale recombination morphology resembles the reionization morphology, but with a delay corresponding to the recombination time-scale $t_{\rm rec} \sim$ 40 Myr $\Delta^{-1} (1-x_{\rm HI})^{-1} [(1+z)/20]^{-3}$.  Most recombinations occur in systems with overdensities close to the self-shielding threshold, which increases from $\Delta_{\rm ss} \sim 5 \rightarrow 100$ over the redshift interval $z\sim 15 \rightarrow 5$ (e.g. Fig. \ref{fig:D_evo}).  Such an increase in $\Delta_{\rm ss}$ offsets the $t_{\rm rec}\propto (1+z)^{-3}$ growth of the recombination time of the mean Universe, such that the recombination time of LLSs remains fairly constant.  This, combined with the growth of cosmic time per redshift 
interval (increase in $dt/dz$ with decreasing redshift), explains why recombinations become more important at lower redshifts during the late stages of reionization.

From Fig. \ref{fig:light_box}, we also see that the combined reionization-era impact of the evolution of sources and sinks is much more potent than either effect on its own.  This is due to the fact that both preferentially affect large HII regions, late in reionization.  Larger regions generally have a higher UVB due to a larger $\lambda_{\rm mfp}$, driving ionization fronts into dense, rapidly recombining systems. Furthermore, the large HII regions are the ones whose centers were the first to ionize, with enough time passing for even modest overdensities to recombine.  This passage of time also depletes the gas reservoir available for star formation through Jeans filtering.  Both effects slow the growth of large HII regions, driving them into a ``photon-starved'' or ``recombination-limited'' regime. We quantify this further below.

\subsection{Reionization History}
\label{sec:xHI_evo}

\begin{figure}
\vspace{+0\baselineskip}
{
\includegraphics[width=0.45\textwidth]{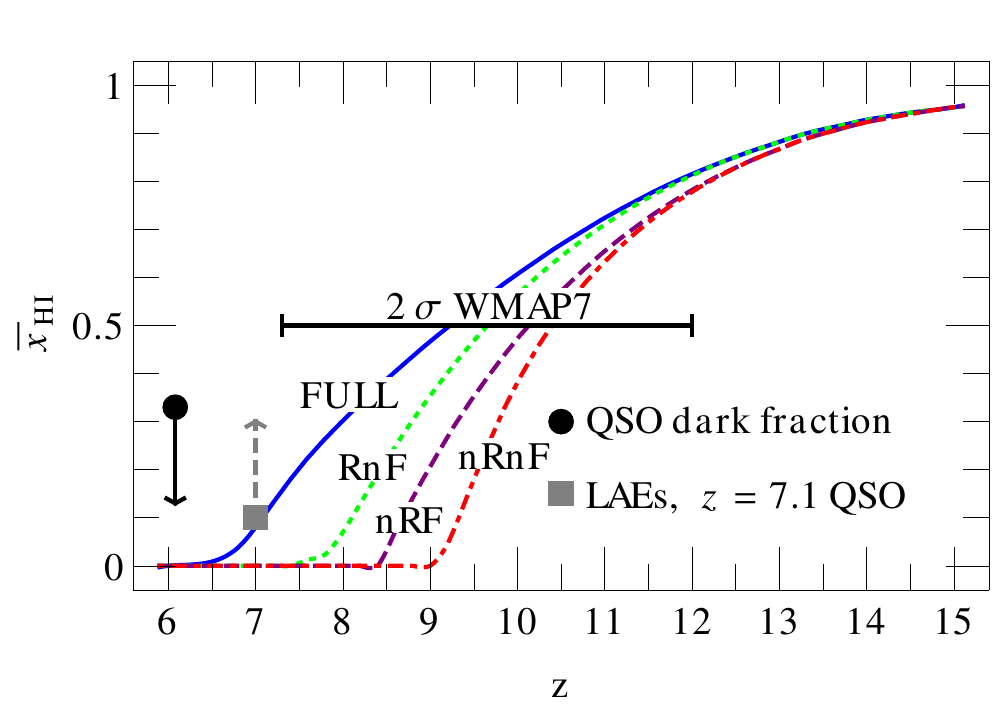}
}
\caption{Evolution of the volume-weighted global neutral fraction, $\bar{x}_{\rm HI}$, with different models. \textit{Solid}: \textbf{FULL}. \textit{Dotted}: \textbf{RnF}. \textit{Dashed}: \textbf{nRF}. \textit{Dot-dashed}: \textbf{nRnF}. For comparison, we show the 2-$\sigma$ constraints from the seven-year WMAP measurement of $\tau_{\rm e}$ (\citealt{WMAP11}; we translate these integral constraints to limits on $\avenf=0.5$ using suites of reionization histories from \citealt{MMS12}).  We also denote the strict upper limit at $z\approx6$ from the dark fraction in QSO spectra \citep{MMF11}, and the two recent (somewhat qualitative) lower limits at $z\approx7$ suggested by (i) observations of ULAS J1120+0641 \citep{BHWH11} and (ii) the fall in the $\text{Ly}\alpha$ emitter fraction among LBGs (e.g. \citealt{DMW11, Pentericci11, BH13}).
\label{fig:xHI_evo}
}
\vspace{-1\baselineskip}
\end{figure}

In Figure \ref{fig:xHI_evo} we compare the evolution of the global volume-averaged
neutral fraction $\bar{x}_{\rm HI}$ in different reionization models\footnote{Note that reionization in our \textbf{nRnF} model evolves quicker than in previous, comparable semi-numerical predictions.  This is because we remove the former crude LLSs modeling which imposed a homogeneous (usually redshift-independent) mean free path, $R_{\rm max}=\lambda_{\rm mfp, HII}$. 
By not including even this crude treatment of LLSs, our \textbf{nRnF} model is more similar to most analytic estimates and large-scale radiative transfer simulations, than it is to previous semi-numerical ones.}.
As mentioned before, both recombinations and UVB feedback delay reionization; the former by depleting the photon budget for growing the HII regions, and the later by decreasing the star-formation rate.   We stress again that the fiducial choice of $\xi=30$ is just a rough guess; reionization histories can be shifted later/earlier by decreasing/increasing $\xi$.

From Fig. \ref{fig:xHI_evo} we also see that the reionization delay from recombinations is more significant than the one from UVB feedback on sources.  By decreasing the effective ionizing efficiency (see Section \ref{sec:rec}), recombinations delay the end of reionization by $\Delta z\sim 1.3$ (\textbf{RnF} vs \textbf{nRnF}).  The analogous delay due to feedback on sources is $\Delta z \sim 0.7$ (\textbf{nRF} vs \textbf{nRnF}). The total effect of both recombinations and feedback on sources is to delay the completion of reionization by $\Delta z \sim 2.5$ (\textbf{FULL} vs \textbf{nRnF}).
  This delay is greater than the combined individual delays from either effect, as both preferentially impact the same, late-stage large HII regions, accelerating their transition to a recombination-limited regime.  Since the impact is strongest on the end stages of (and following) reionization, $\tau_{\rm e}$ is only mildly affected, with the \textbf{FULL} (\textbf{nRnF}) model having $\tau_{\rm e}$= 0.07 (0.08).


\subsection{Reionization Morphology}
\label{sec:morph}

\begin{figure*}
\vspace{+0\baselineskip}
{
\includegraphics[width=0.33\textwidth]{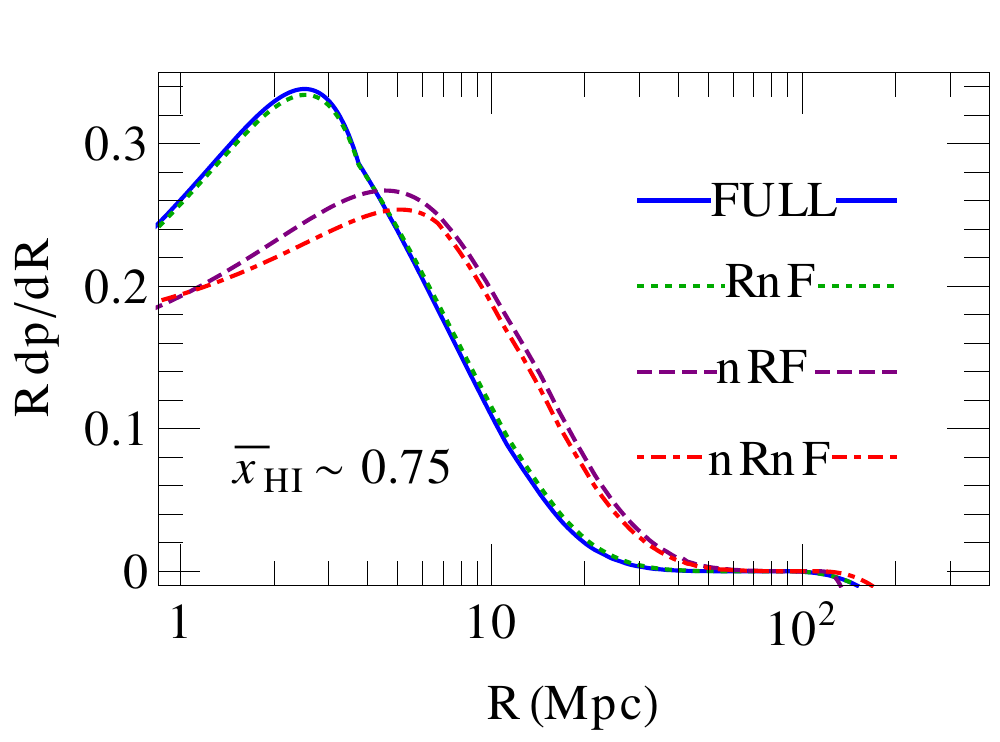} 
\includegraphics[width=0.33\textwidth]{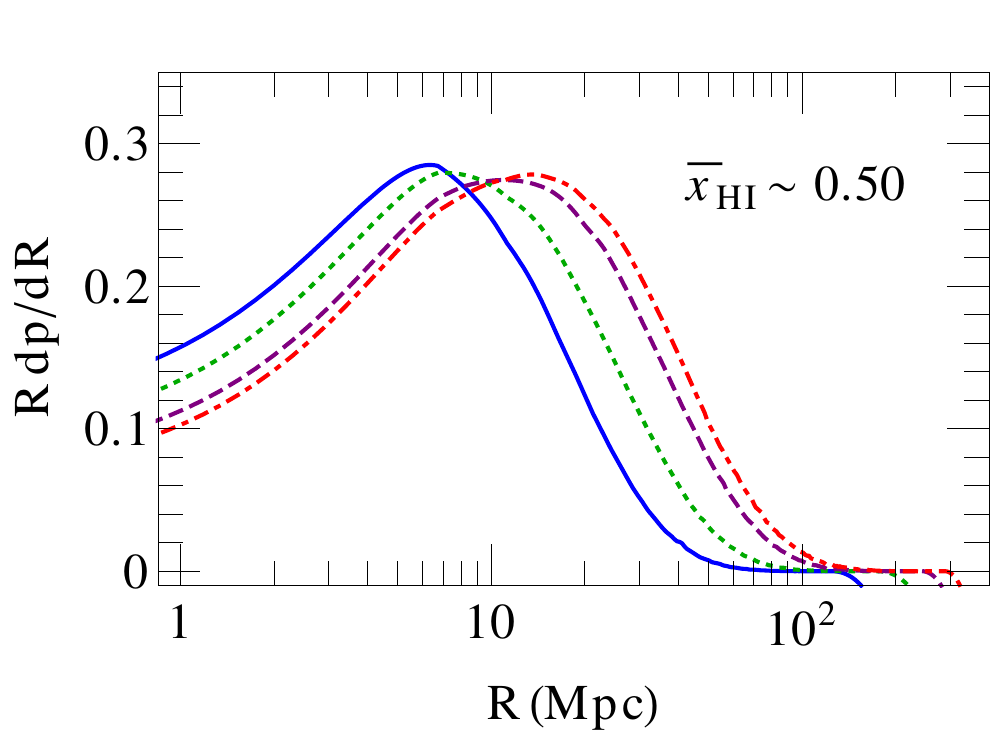} 
\includegraphics[width=0.33\textwidth]{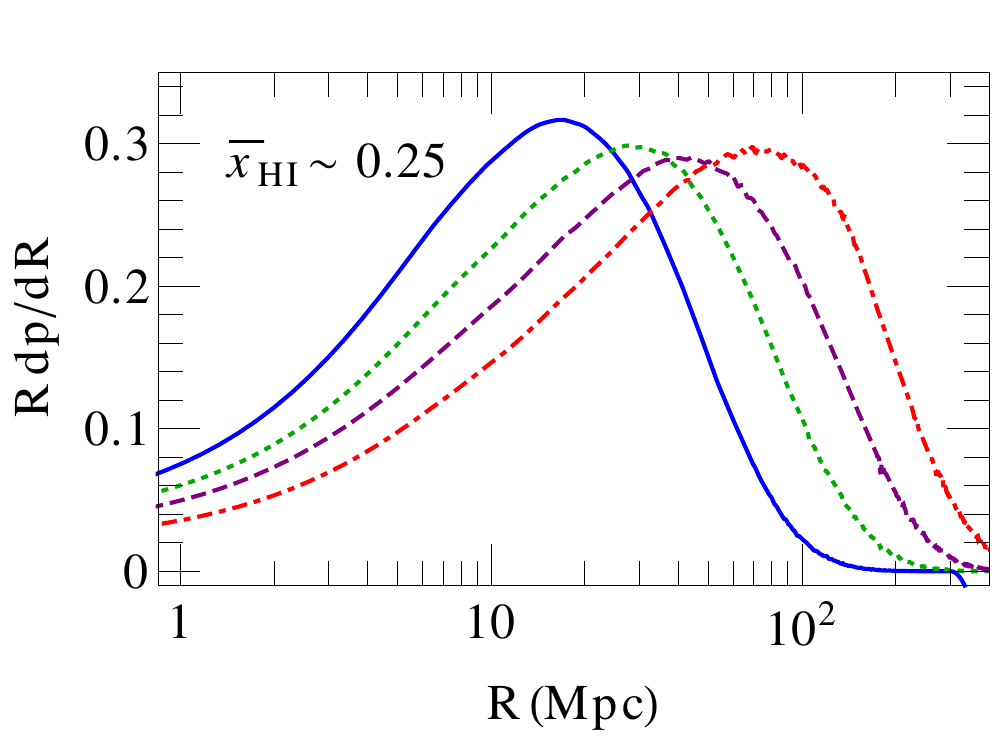}
\includegraphics[width=0.33\textwidth]{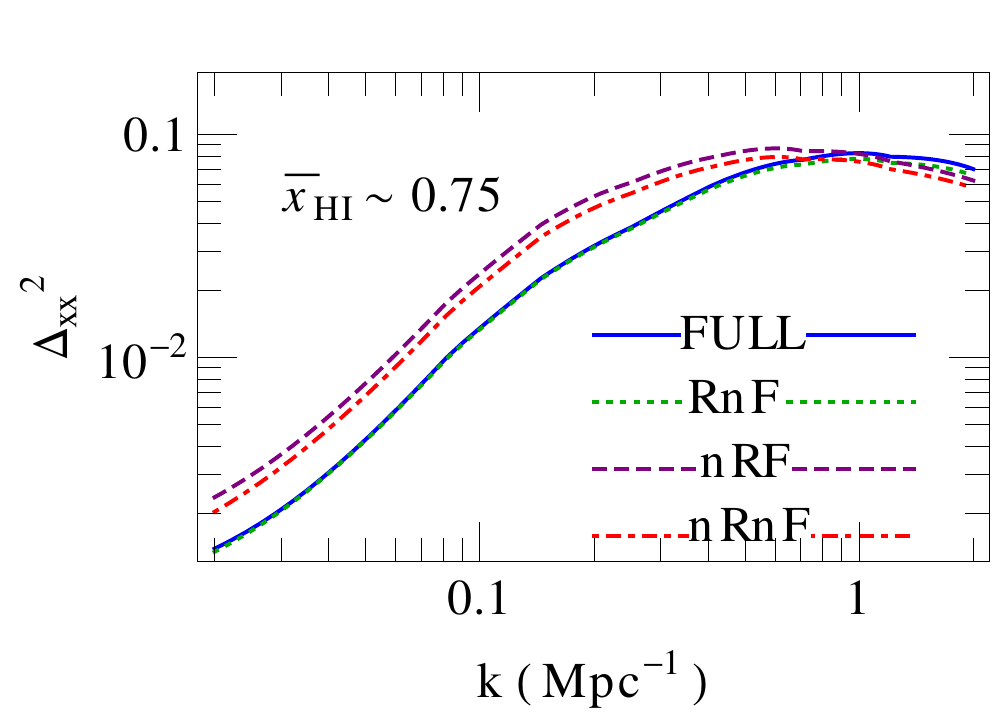} 
\includegraphics[width=0.33\textwidth]{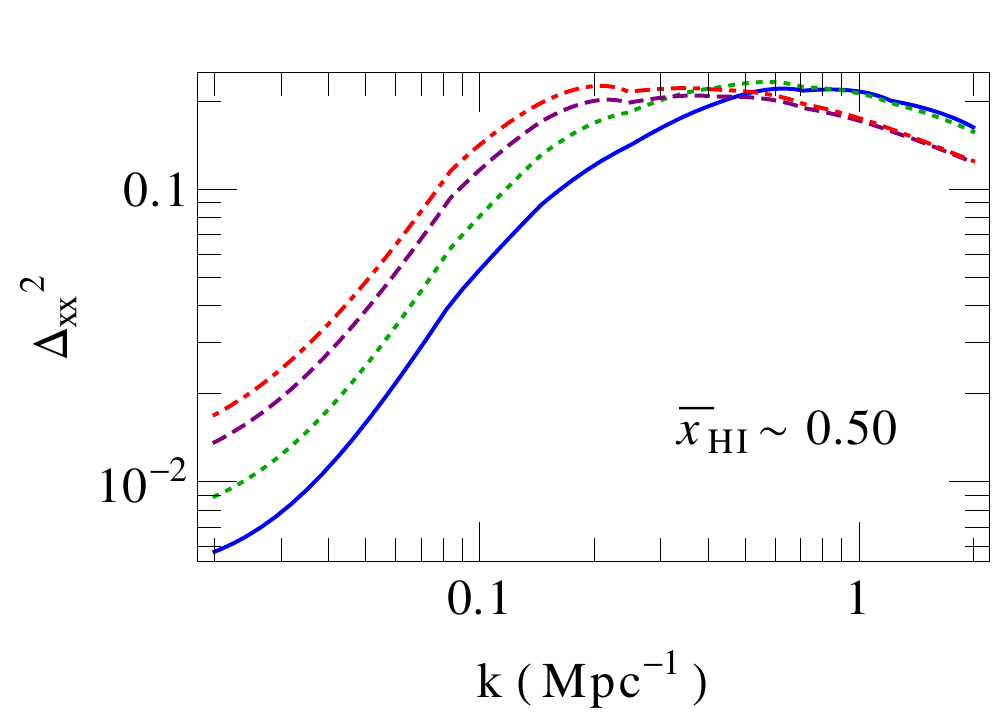} 
\includegraphics[width=0.33\textwidth]{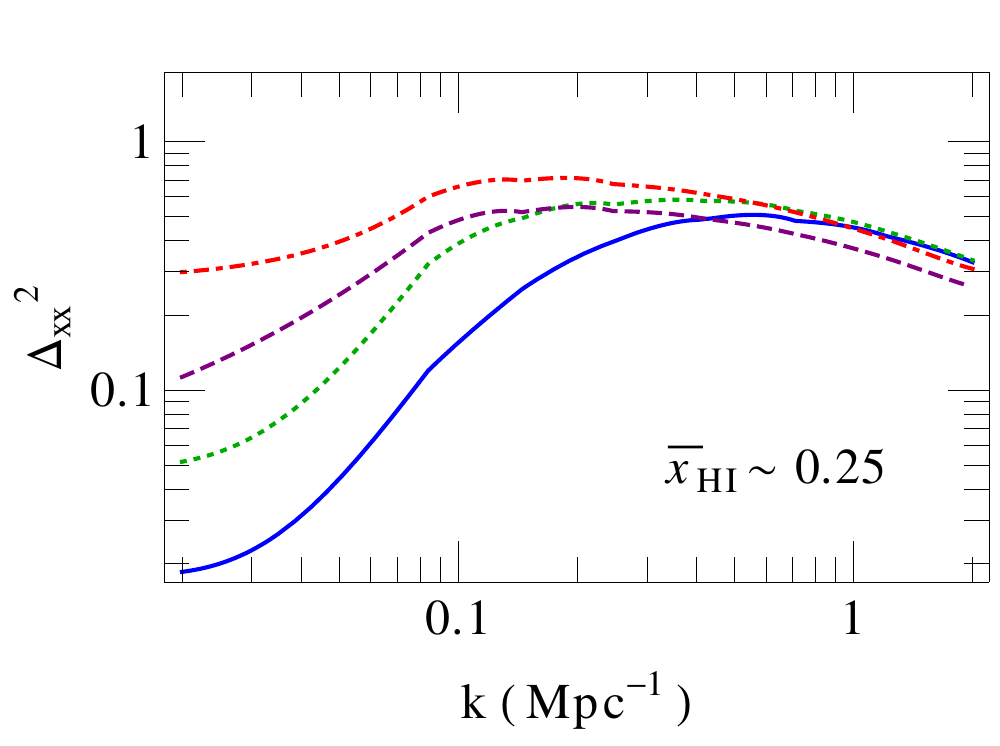}
\includegraphics[width=0.33\textwidth]{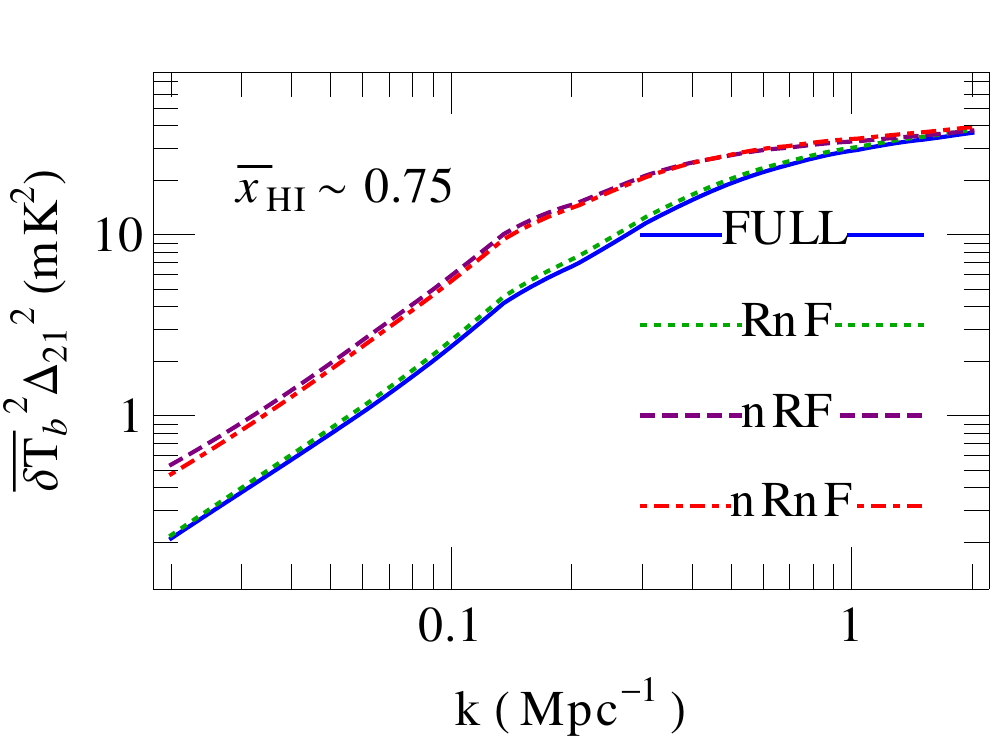} 
\includegraphics[width=0.33\textwidth]{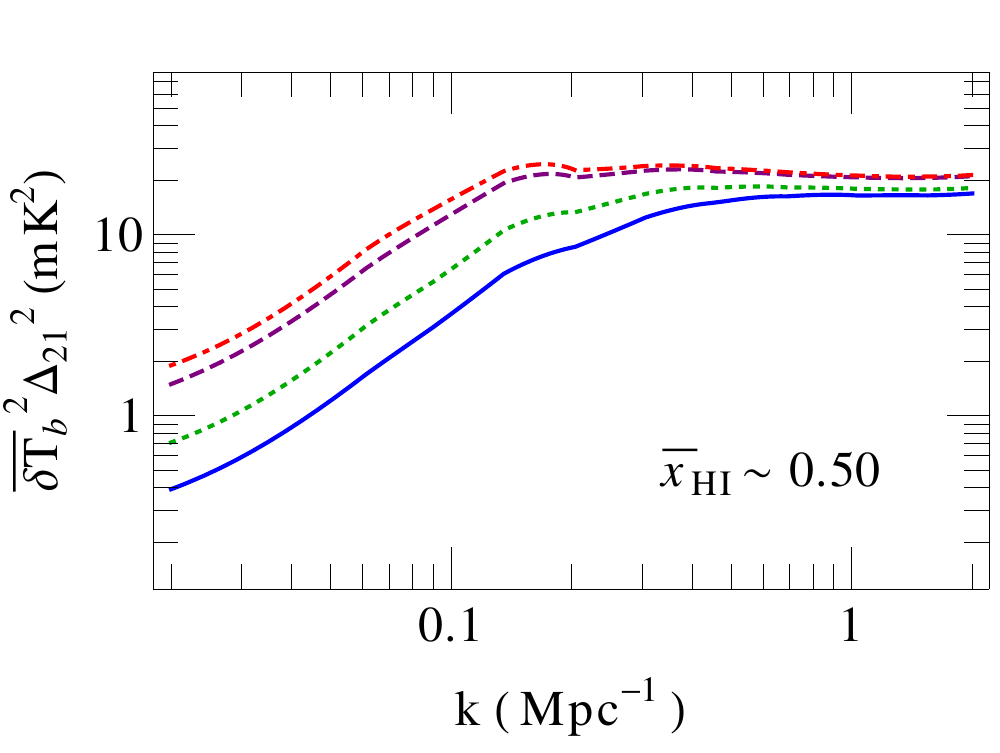} 
\includegraphics[width=0.33\textwidth]{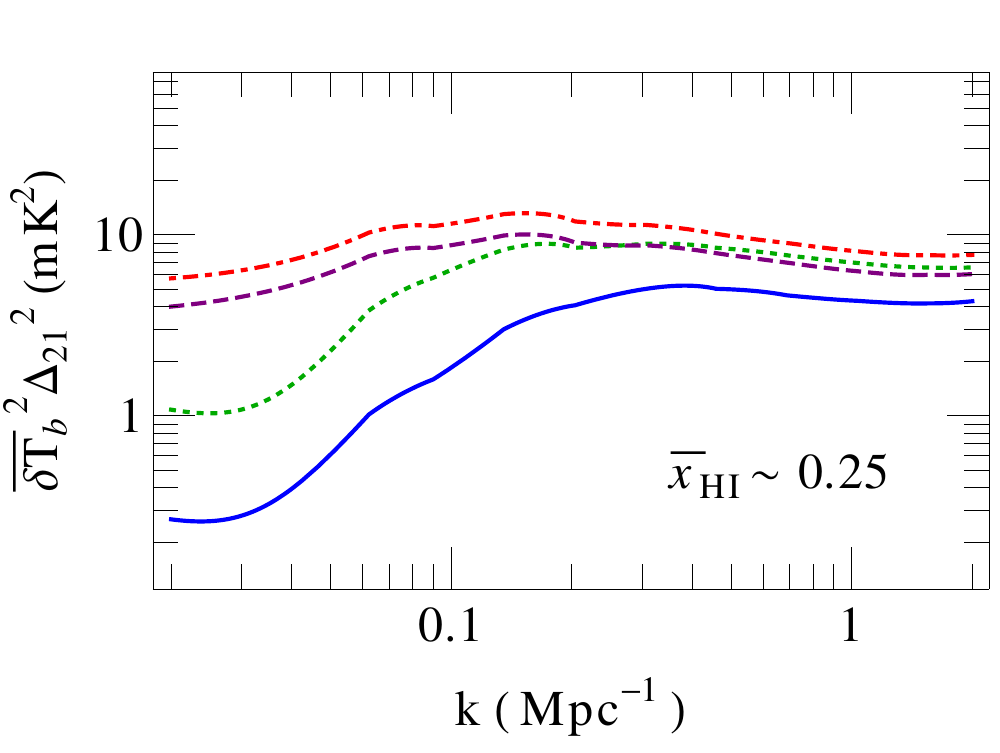}
}
\caption{Size distributions of the HII regions ({\it top panels}), power spectra of the ionization field ({\it middle panels}) and of the 21cm field ({\it bottom panels}) at $\bar{x}_{\rm HI}\approx 0.75$ (\textit{left}), $\bar{x}_{\rm HI}\approx 0.5$ (\textit{middle}) and $\bar{x}_{\rm HI}\approx 0.25$ (\textit{right}). \textit{Solid}: \textbf{FULL}. \textit{Dotted}: \textbf{RnF}. \textit{Dashed}: \textbf{nRF}. \textit{Dot-dashed}: \textbf{nRnF}. Note that the y-axis in the three panels spans different ranges.
\label{fig:pow_spec}
}
\vspace{-1\baselineskip}
\end{figure*}

We now proceed to quantify the impact of recombinations on the morphology of reionization. Understanding the morphology of reionization is important in interpreting almost all reionization observables (e.g. \citealt{Lidz07, McQuinn08, MF08damp, Mesinger10, SMH13, DMW11, DMF11, MMS12}).  Furthermore, current 21cm interferometers, such as Low Frequency Array (LOFAR; \citealt{vanHaarlem13})\footnote{http://www.lofar.org/},  Murchison Wide Field Array (MWA; \citealt{Tingay12})\footnote{http://www.mwatelescope.org/}, and the Precision Array for Probing the Epoch of Reionization (PAPER; \citealt{Parsons10})\footnote{http://eor.berkeley.edu}, should provide a statistical measurement of large-scale ($k\sim0.1$ Mpc$^{-1}$) reionization morphology in the next couple of years.

In the top panels of Fig. \ref{fig:pow_spec} we show the size distributions of the HII regions, calculated according to the procedure in \citet{MF07}: randomly choosing an ionized cell and tabulating the distance to the HII region edge along a randomly chosen direction.  In the middle panels we show the ionization power spectrum $\Delta^2_{\rm xx} \equiv k^3/(2\pi^2 V) ~ \langle|\delta_{\rm xx}|^2\rangle_k$, with $\delta_{\rm xx}=x_{\rm HI}/\avenf - 1$.  Panels correspond to different stages of reionization: $\bar{x}_{\rm HI}\approx 0.75, 0.50, 0.25$ ({\it left to right}). We compare our fiducial model \textbf{FULL} (\textit{solid}) with the runs \textbf{RnF} (\textit{dotted}), \textbf{nRF} (\textit{dashed}) and \textbf{nRnF} (\textit{dot-dashed}).
To facilitate direct comparisons to upcoming 21cm interferometric observations, in the bottom panels we plot the corresponding 21cm power spectrum, defined as $k^3/(2\pi^2 V) ~ \bar{\delT}(z)^2 ~ \langle|\delta_{\rm 21}({\bf k}, z)|^2\rangle_k$, where $\delta_{21}({\bf x}, z) \equiv \delT({\bf x}, z)/ \bar{\delT}(z) - 1$, and the brightness temperature offset from the CMB is computed assuming that the spin temperature is much greater than the CMB temperature:
\begin{align}
\label{eq:delT}
 \delT(\nu) = &27 \nf \Delta \left(\frac{H}{dv/dr + H}\right) \sqrt{\frac{1+z}{10} \frac{0.15}{\Omega_{\rm m} h^2}} \left( \frac{\Omega_{\rm b} h^2}{0.023} \right) {\rm mK},
\end{align}
\noindent where  $H(z)$ is the Hubble parameter, $dv/dr$ is the comoving gradient of the line of sight component of the comoving velocity, and all quantities are evaluated at redshift $z=\nu_0/\nu - 1$.


As discussed above, UVB feedback on sources does result in a more uniform distribution of HII regions, thus suppressing the power spectra on large scales (e.g. \citealt{QLZD07}).  However the suppression is at level of $10$s of per cent around $k\sim0.1$ Mpc$^{-1}$ at $\avenf\sim 0.25$ \citep{SM13b}.  

Instead, recombinations have a much stronger impact.  The scale of the suppression, $\gsim 10\text{ Mpc}$, (when bubbles freeze-out from recombination limited growth) is roughly consistent with analytic predictions \citep{FO05}.  However, the impact of recombinations is significant much earlier than the recombination-limited regime ($\avenf \approx 0.1$; see Fig. \ref{fig:J21_evo} and \citealt{FO05}).  This is due to the fact that recombinations start slowing down the growth of HII regions well in advance of their Str{\"o}mgren limit (see the evolution of $\xi_{\rm eff}/\xi$ in Fig. \ref{fig:rec}).  Indeed, we find that the requirement that the local region of size $R$ has enough time-integrated photons to overcome the time-integrated recombinations (our eq. \ref{eq:ion_crit_coll}) is more restrictive than the {\it instantaneous}, $\lambda_{\rm mfp, HII} \geq R$ criterion, used in previous works (e.g. \citealt{FO05, CHR09}).

The effect is dramatic.  Recombinations decrease the typical HII region size by factors of $\sim$2--3 {\it throughout} reionization. During the second half of reionization, recombinations suppress the power spectra by $\gsim50$--100\% (\textbf{RnF} vs \textbf{nRnF}) on the $k\sim0.1$ Mpc$^{-1}$ scales relevant for current 21cm interferometers (e.g. \citealt{Pober13, ME-WH13}).  
For the \textbf{FULL} model which also includes UVB feedback on sources, this suppression increases to a factor of $\gsim 2$--4.\footnote{The dearth of large-bubbles increases the cross-correlation between the ionization and density fields, which has a negative contribution to the 21cm power spectrum.  This generally results in a somewhat larger suppression of the 21cm power spectra than the ionization power spectra.}

This dramatic suppression of large-scale power from recombinations qualitatively impacts the shape of the ionization power spectrum.  Neglecting recombinations, large-scale, radiative transfer simulations (e.g. \citealt{QLZD07, ZMQT11, Friedrich11}) predict a relatively flat or mildly decreasing power spectrum in the relevant range $k\sim0.1\rightarrow0.3$, during the late stages of reionization, consistent with our \textbf{nRnF} model (c.f. \citealt{ZMQT11}).  The dramatic suppression of large-scale power caused by recombinations instead results in a strongly-increasing power spectrum in this range.  The amplitude and slope of the power spectrum around $k\sim0.1$ Mpc$^{-1}$ are fundamental observables of the first generation 21cm interferometers (e.g. \citealt{Lidz08}).  {\it Recombinations strongly affect both.}

Our model explicitly tracks the sub-grid mass fraction of neutral gas inside the HII regions, finding it to be a few percent on average, consistent with post-reionization measurements (e.g. \citealt{WGP05}).  Instead if the (mass-weighted) neutral fraction of the entire $\sim1$ Mpc simulation cell was set to unity, the reionization morphology could be noticeably changed on small-scales \citep{CHR09, CMMF11}.  Indeed, the procedure outlined in \citealt{CHR09} assumes that an entire $\sim 1.4$ cMpc cell is neutral, if it begins to self-shield.  They find that this large reservoir of neutral gas can drive up the small-scale 21cm power. This is contrary to our results, which are driven by the sub-grid density and ionization distributions.

\subsection{Additional Properties of the Ionized IGM}
\label{sec:obs_reion}

One of the benefits of our formalism is that it can track IGM evolution into the late stages of reionization, and even {\it somewhat} afterwards\footnote{We caution against over-interpreting our results at late times for several reasons. Firstly, our analysis assumes an infinite speed of light, and this assumption becomes increasingly worse as reionization completes, and the mean free path grows. Since we do not integrate along the light cone, we expect our model to break down when the emissivity evolves significantly during the light-crossing time of $\lambda_{\rm mfp}$, i.e $\varepsilon/\dot{\varepsilon}\lesssim\lambda_{\rm mfp}/c$. Relatedly, we ignore the redshifting of ionizing photons that likely becomes important for such soft spectra as the mean free path increases.  Finally, as already mentioned, the MHR density distribution breaks down at very high densities, meaning that our modeling becomes inaccurate when $\Delta_{\rm ss}$ becomes very large.}.
This allows us to test our model against IGM observations at redshifts $z\lsim6$, before the \lya\ forest saturates.  Here we show the evolution of the mean free path, the ionizing background and the ionizing emissivity.  It is useful to recall (eq. \ref{eq:self_cons_J}) that inside ionized regions (and post reionization), these quantities approximately scale as: $\Gamma_{\rm HII} \propto \lambda_{\rm mfp} \varepsilon$.  The ionizing background can be measured from the \lya\ forest of high-$z$ quasars (with the a-priory assumption that reionization has completed; \citealt{Mesinger10}), while $\lambda_{\rm mfp}$ can be either measured or estimated with radiative transfer simulations.  These measurements become increasingly uncertain at $z\gsim4$ (e.g. \citealt{BH07,QOF11, KF12}).

\subsubsection{Ionizing emissivity and recombination rate}
\label{sec:em_evo}

\begin{figure}
\vspace{+0\baselineskip}
{
\includegraphics[width=0.45\textwidth]{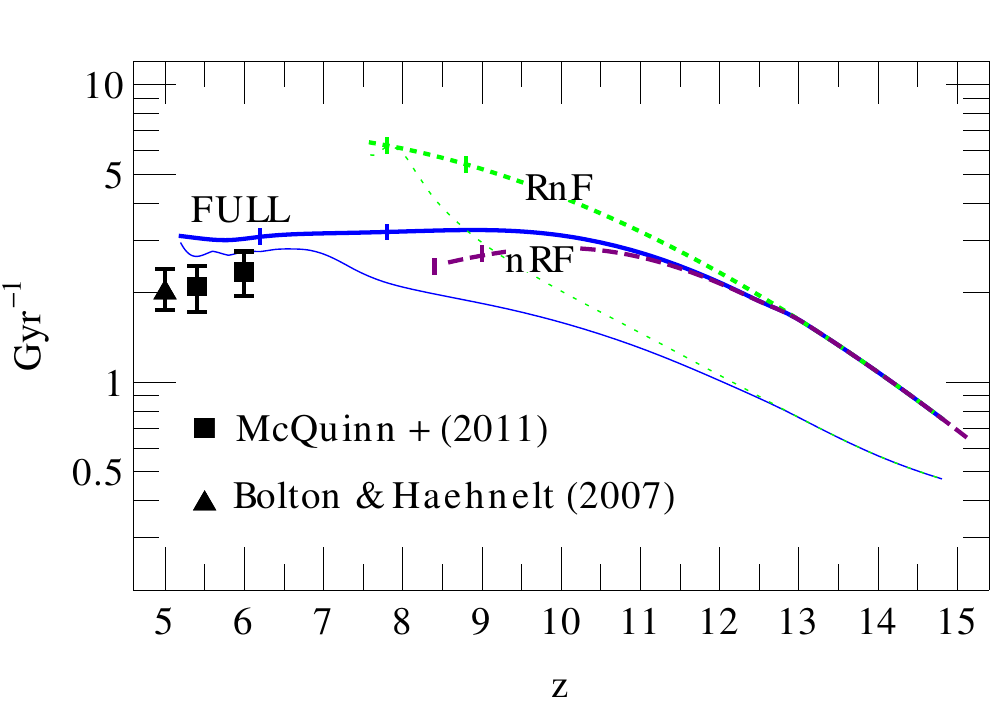}
}
\caption{Evolution of the average emissivity (\textit{thick}) and recombination rate per baryon (\textit{thin}) with different models. \textit{Solid}: \textbf{FULL}. \textit{Dotted}: \textbf{RnF}. \textit{Dashed}: \textbf{nRF}. The vertical ticks correspond to $\avenf =0.2$ and $\avenf =10^{-2}$. For comparison we show the emissivity constraints inferred from the $\text{Ly }\alpha$ forest at $z \lsim 6$ \citep{BH07,QOF11}.
\label{fig:em_evo}
}
\vspace{-1\baselineskip}
\end{figure}

In Figure \ref{fig:em_evo} we show the evolution of the ionizing photons emissivity $\varepsilon\equiv\xi df_{\rm coll}/dt$ (\textit{thick}) and the recombination rate per baryon (\textit{thin}) for the models \textbf{FULL} (\textit{solid}), \textbf{RnF} (\textit{dotted}) and \textbf{nRF} (\textit{dashed}); the emissivity in the model \textbf{nRnF} is the same as in the model \textbf{RnF}. We plot the curves until $\lambda_{\rm mfp}$ exceeds the box size. The vertical ticks corresponds to the redshifts when the volume-averaged neutral fraction is $\avenf =0.2$ and $\avenf=10^{-2}$.  For comparison we show the emissivity constraints inferred from the $\text{Ly}\alpha$ forest \citep{BH07,QOF11}.

At the earliest redshifts, all of the runs have the same emissivity, by construction.  As time passes, the runs which include UVB feedback on star-formation have a reduced emissivity, consistent with $z\lsim6$ observations.  Furthermore, the emissivity is somewhat (10s of per cent) higher in the run \textbf{FULL} than in the run \textbf{nRF}: if we neglect recombinations, $\bar{\Gamma}_{\rm HII}$ increases and reionization occurs earlier, resulting in stronger UVB feedback.  Overall, UVB feedback decreases the emissivity by a factor of $\sim2$ by the end of reionization.  This is relatively modest when compared to the corresponding impact on $\Gamma_{\rm HII}$, which can scale with the emissivity as a power law, in the late (recombination-limited) stages of reionization and afterwards (\citealt{QOF11}; see also below).

Runs which include recombinations reach a recombination-limited regime at $\avenf\sim0.1$ (the short, vertical ticks on the curves demarcate $\avenf=0.2$, 0.01), consistent with analytic predictions \citep{FO05}.  Runs \textbf{RnF} (\textbf{FULL}) result in $\approx$0.9 (1) recombinations per baryon by the end of reionization.   It is important to note that an {\it extended} regime of a constant emissivity and recombination rate only occurs in the \textbf{FULL} model, with the {\it combined effort} of the evolutions of sources and sinks slowing down reionization sufficiently.  As we shall see below, this regulates the rise in the $\lambda_{\rm mfp}$ and $\Gamma_{\rm HII}$ in the final stages of reionization.

We stress again that our runs are only ``tuned'' via the fiducial choice of ionizing efficiency $\xi=30$ so as to match the observed $\tau_{\rm e}$.  If one plays with this free parameter, one can shift the emissivities in Figure \ref{fig:em_evo} up or down.  We test this explicitly with a new \textbf{RnF} run, but with $\xi=15$ (not shown in the figure for the sake of clarity).  The emissivity in this run is approximately consistent with observational estimates at $z\sim5$--6, without appealing to UVB feedback on sources.  However, in this case reionization happens relatively late, with a corresponding $\tau_{\rm e}=0.058$, inconsistent with WMAP at $\approx$2--3 $\sigma$.

\subsubsection{The mean free path and the UVB}
\label{sec:MFP_evo}

\begin{figure}
\vspace{+0\baselineskip}
{
\includegraphics[width=0.45\textwidth]{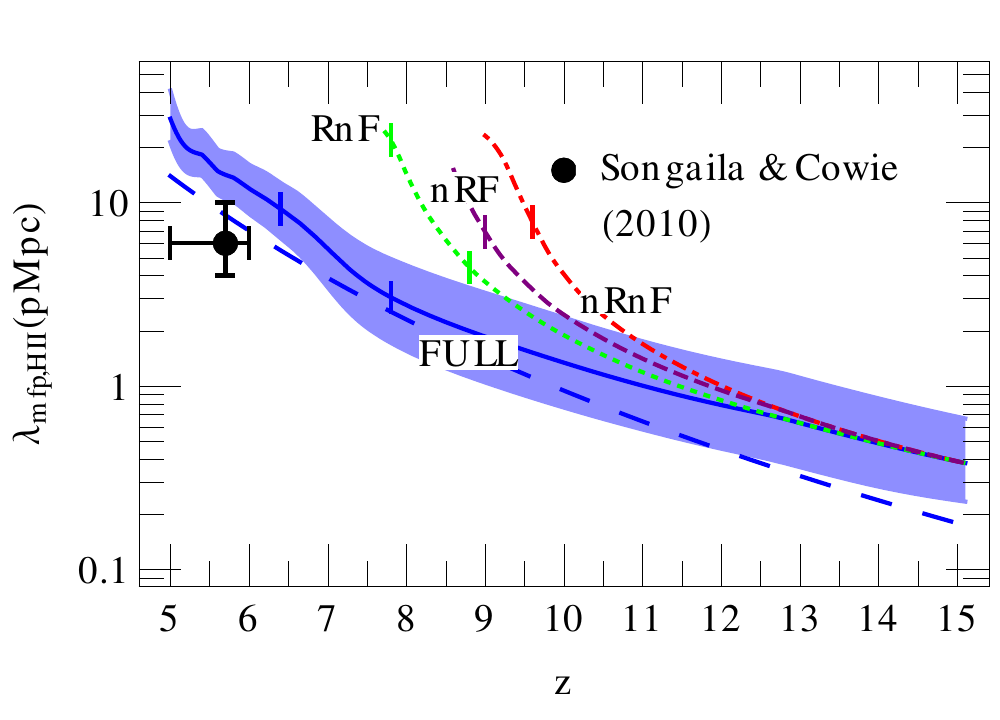}
}
\vspace{-1\baselineskip}
\caption{Evolution of the spatially-averaged mean free path inside the ionized IGM, $\bar{\lambda}_{\rm mfp, HII}$, in proper units. \textit{Solid}: \textbf{FULL}; the shaded region corresponds to the 1-$\sigma$ spatial scatter. \textit{Dotted}: \textbf{RnF}. \textit{Dashed}: \textbf{nRF}. \textit{Dot-dashed}: \textbf{nRnF}. \textit{Long-dashed}: analytic formula provided by \citet{SC10}; for comparison we show observations by the same authors. The vertical ticks corresponds to the redshift when $\avenf =0.2$ and $\avenf =10^{-2}$.
\label{fig:mfp_evo}
}
\vspace{-1\baselineskip}
\end{figure}

In Figure \ref{fig:mfp_evo} we compare the evolution of $\lambda_{\rm mfp, HII}$ in our runs.  Only the \textbf{FULL} run is consistent with observations at $z\sim 6$, as well as the high redshift extrapolation of the empirically-calibrated formula provided in \citet{SC10}
\begin{equation}
\lambda_{\rm mfp}=50\left(\frac{1+z}{4.5}\right)^{-4.44}\text{ pMpc} ~ ,
\end{equation}
 which approximates the CDDF as a power law.



\begin{figure}
\vspace{+0\baselineskip}
{
\includegraphics[width=0.5\textwidth]{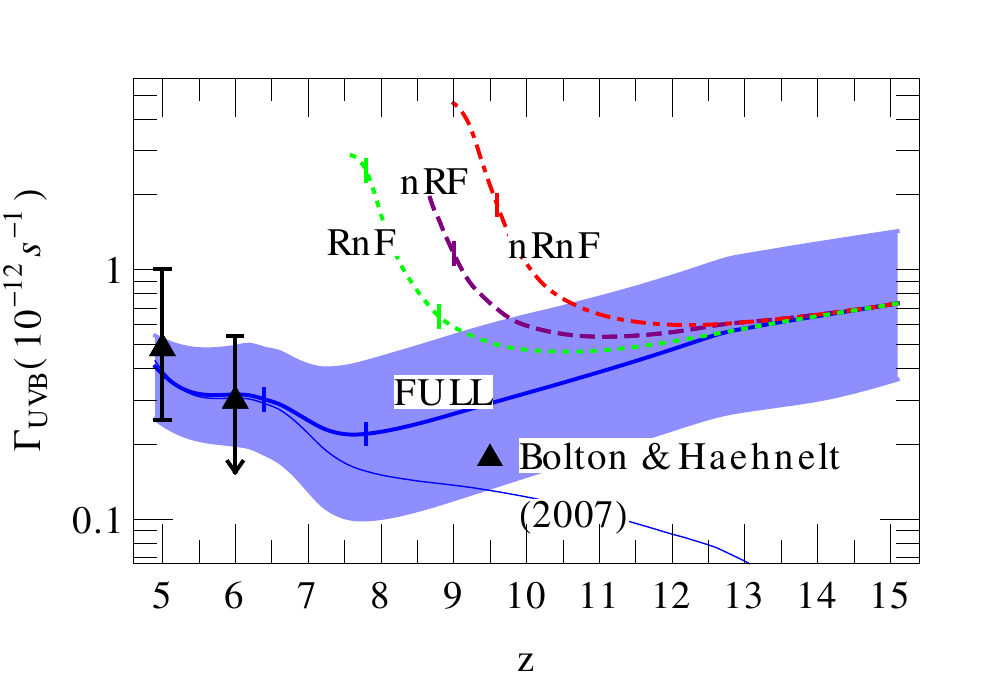}
}
\caption{Evolution of the average photoionization rate $\bar{\Gamma}_{\rm HII}$, in ionized regions.  \textit{Solid}: \textbf{FULL}; the shaded region corresponds to the spread among HII regions. \textit{Dotted}: \textbf{RnF}. \textit{Dashed}: \textbf{nRF}. \textit{Dot-dashed}: \textbf{nRnF}. 
The vertical ticks correspond to $\avenf =0.2$ and $\avenf =10^{-2}$.
 The thin solid curve shows instead the volume-averaged photoionization rate, assuming a negligible contribution from inside neutral patches: $\langle \Gamma \rangle_V \approx [1-\langle x_{\rm HI}\rangle_V] \bar{\Gamma}_{\rm HII}$.
 For comparison we also show observational estimates from the $\text{Ly}\alpha$ forest \citep{BH07}. 
\label{fig:J21_evo}
}
\vspace{-1\baselineskip}
\end{figure}

The same trend can be seen in Fig. \ref{fig:J21_evo}, where we plot the evolution of  $\bar{\Gamma}_{\rm HII}$, {\it averaged over HII regions}.
  Namely, in all runs except \textbf{FULL}, the mean free path and UVB grow too rapidly, too early. 

In order to avoid rapid, early evolution in the UVB and mean free path, one needs {\it both}: (i) recombinations; {\it and} (ii) a fairly low emissivity, $\varepsilon \lsim 3$ Gyr$^{-1}$. In our fiducial model, (ii) is assured by UVB feedback on star formation, which also results in an extended reionization history.  As mentioned previously, in the absence of UVB feedback, a low emissivity could also result from a smaller ionizing efficiency, $\xi\lsim15$; however in this case reionization would occur rapidly and late, and be only marginally consistent with WMAP observations.  Similarly, a low emissivity could also be obtained by increasing $M_{\rm cool}$, physically corresponding to reionization driven by more-massive sources which generate ionizing photons more efficiently.  However in this case, reionization again occurs very late, and with the added complication of an emissivity which, although low at $z\sim6$, is evolving too rapidly to be consistent with observations of the \lya\ forest (\citealt{
Miralda03}; Fig. 12 in \citealt{MMS12}).

\subsubsection{Clumping Factor}
\label{sec:clumping}

\begin{figure}
\vspace{+0\baselineskip}
{
\includegraphics[width=0.45\textwidth]{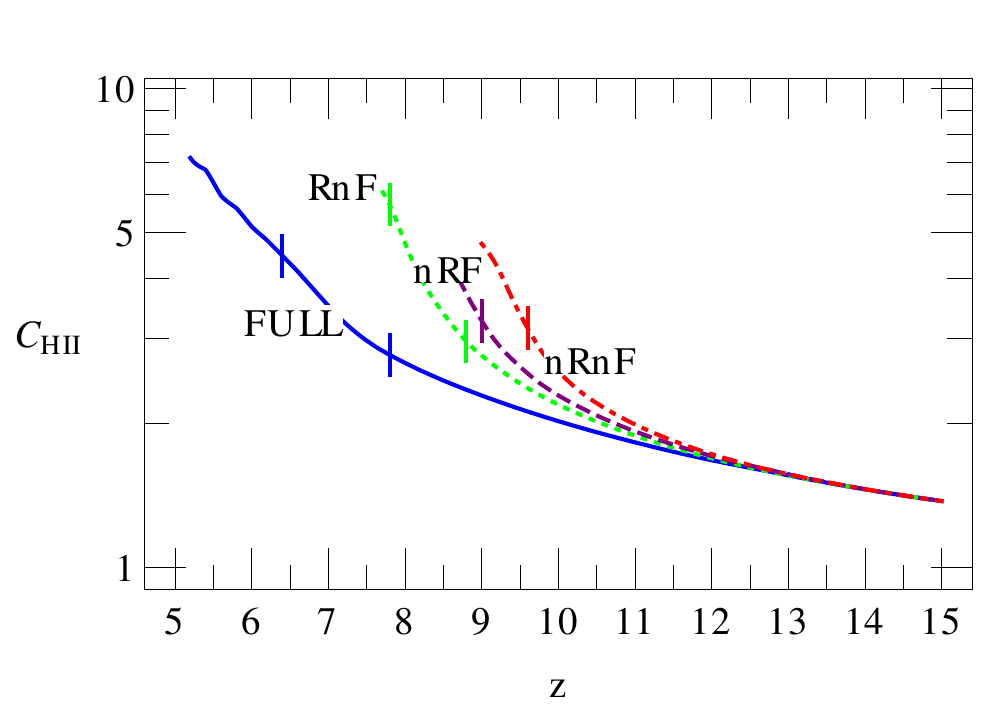}
}
\caption{Evolution of the ionized gas clumping factor. \textit{Solid}: \textbf{FULL}.  \textit{Dotted}: \textbf{RnF}. \textit{Dashed}: \textbf{nRF}. \textit{Dot-dashed}: \textbf{nRnF}. The vertical ticks correspond to $\bar{x}_{\rm HI}=0.2$ and $\bar{x}_{\rm HI}=10^{-2}$.
\label{fig:CF_evo}
}
\vspace{-1\baselineskip}
\end{figure}

\begin{figure}
\vspace{+0\baselineskip}
{
\includegraphics[width=0.5\textwidth]{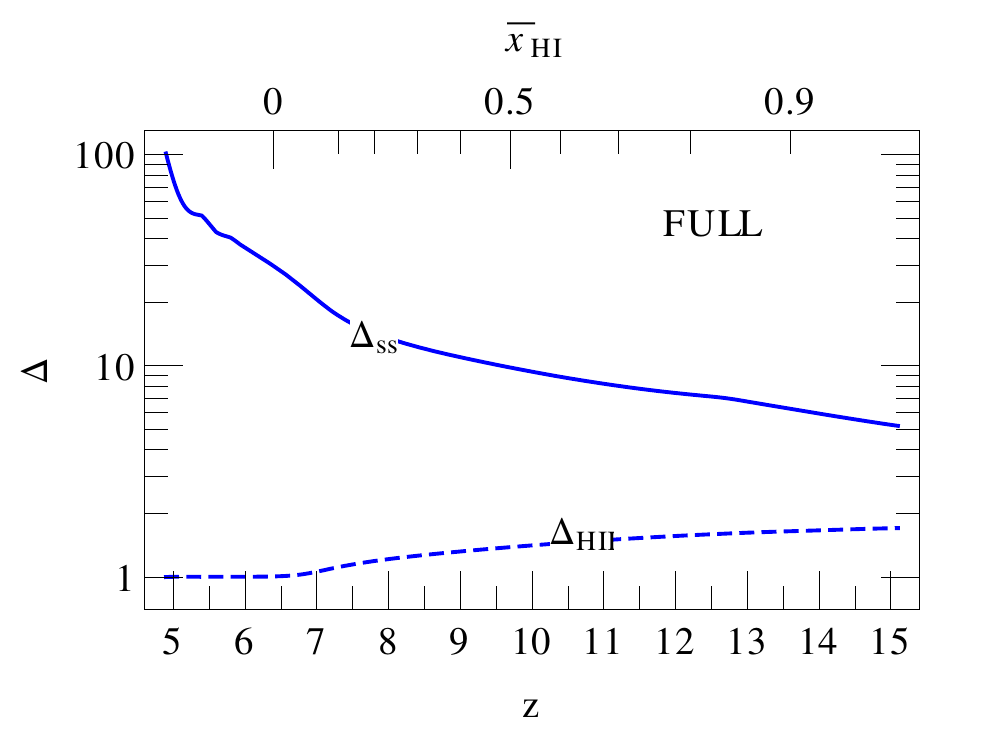}
}
\caption{Evolution of the relevant overdensity thresholds in HII regions for the run \textbf{FULL}. \textit{Solid}: overdensity for efficient self-shielding (eq. \ref{eq:D_ss}). \textit{Dashed}: average, large-scale overdensity in HII regions.
\label{fig:D_evo}
}
\vspace{-1\baselineskip}
\end{figure}

In analytic models, recombinations are often parameterized in terms of a ``clumping factor'' of the ionized gas.  This quantity encodes the integral over the ionization structure of the gas, facilitating analytic estimates of the global recombination rate using a single parameter.  There are several ways of defining the clumping factor (see, e.g. \citealt{FOOD12}).  Here we use the following simple definition: $C_{\rm HII}\equiv\langle n_{\rm HII}^2\rangle / \langle n_{\rm H}\rangle^2$. In Figure \ref{fig:CF_evo} we show the evolution of $C_{\rm HII}$ in the runs \textbf{FULL} (\textit{solid}), \textbf{RnF} (\textit{dotted}), \textbf{nRF} (\textit{dashed}) and \textbf{nRnF} (\textit{dot-dashed}).

The evolution of $C_{\rm HII}$ can be understood looking at the overdensity threshold for self-shielding, $\Delta_{\rm ss}$ (eq. \ref{eq:D_ss}), and the large-scale mean overdensity in HII regions, $\Delta_{\rm HII}$. In Figure \ref{fig:D_evo} we show the evolution of $\Delta_{\rm ss}$ (\textit{solid}) and $\Delta_{\rm HII}$ (\textit{dashed}) in the run \textbf{FULL}. $\Delta_{\rm ss}$ is increasing with time, driven by its dependence on redshift and by the growth of $\Gamma_{\rm HII}$ in the late stages of reionization [$\Delta_{\rm ss}\propto\left(1+z\right)^{-3}\Gamma_{\rm HII}^{2/3}$].  Physically, this corresponds to ionization fronts penetrating into increasingly overdense regions of the IGM (with higher recombination rates), thereby driving the rapid increase of $C_{\rm HII}$ during the late stages of reionization. At the end of reionization, the value of the clumping factor for the run \textbf{FULL} ($C_{\rm HII}\sim 4$) is consistent with previous works (e.g. \citealt{PSS09, QOF11, SHTS12, 
FOOD12}).

On the other-hand, during the early stages of reionization, HII regions are biased towards the large-scale overdensities which host the first generations of galaxies. Since $C_{\rm HII}\propto \Delta_{\rm HII}^2$, the decreasing bias of the HII regions (fall in $\Delta_{\rm HII}$ with time) balances the rise in $\Delta_{\rm ss}$, resulting in a relatively flat $C_{\rm HII}\approx 1.5$--2 at $z\gsim10$ ($\avenf \gsim 0.5$) (see also, e.g., \citealt{KG13, So13}).

\section{Conclusions}
\label{sec:concl}

We study the role of ionizing photon sinks during reionization by implementing a sub-grid recipe inside large-scale, semi-numeric simulations of reionization.  Building on previous work, we self-consistently model the evolution of both sinks and sources of ionizing photons, using prescriptions calibrated to numerical simulations.  The inhomogeneous recombination rate and emissivity in our model depend on the local density, photo-ionizing background, and reionization history.

We find that both UVB feedback on sources, and recombinations slow the growth of large HII regions, prolonging reionization and resulting in a more uniform ionization morphology.  Although recombinations are more potent, both affects amplify one another as they most strongly affect the same biased regions.  Such regions were the first to ionize, allowing enough time for both recombinations and photoevaporation to take effect.

In our complete model, the end of reionization is delayed by $\Delta z \approx 2.5$.  This delay is sufficient for simple reionization models to match best estimates of the mid and end stages of reionization, as implied by recent observations of the CMB, quasar spectra, and LAEs.

Recombinations are mostly responsible for the slowing and eventual "freeze-out" of large HII regions ($\gsim$10 Mpc).  This results in a dramatic suppression of the large-scale ($k\lsim0.2$ Mpc$^{-1}$) ionization power-spectrum by factors of $\gsim 2$--3, even as early as $\avenf < 0.5$.  Such a dramatic impact on these scales makes recombinations invaluable in interpreting upcoming data from 21cm interferometers.

Furthermore, recombinations are responsible for damping the rapid rise of the mean free path and photo-ionizing background during the late stages of reionization.  The volume-averaged photoionization rate increases by a modest factor of $\sim2$ (as opposed to $\sim5$ ignoring recombinations) during the last, $\avenf<0.2$ stages of reionization.

Our complete model naturally results in an early start, and "photon-starved" end of reionization, as well as a modest, slowly-evolving emissivity (governed by UVB feedback which depletes gas from increasingly massive halos as time progresses).  Although undoubtedly too simplistic, this physical picture ameliorates empirically-motivated claims for a rapid redshift evolution in galaxy properties, like the escape fraction of ionizing photons (e.g. \citealt{HM12, KF12})\footnote{To be more precise, our emissivity estimates are only sensitive to the product on the RHS in eq. (\ref{eq:zeta}); most importantly $f_{\rm esc} f_\ast$, as the dependence on $f_{\rm b}$ is accounted for in our UVB feedback treatment via an evolving $M_{\rm crit}(\textbf{x},z)$.  Our model assumes that this product does not have a halo mass or redshift dependence.
In our model, the emissivity during reionization is dominated by faint galaxies, 1-2 orders of magnitude fainter than current detection limits.  Instead, at lower redshifts ($z\lesssim3$--4), the emissivity from star-forming galaxies is likely dominated by the observed LBGs, which do seem to have somewhat lower values of $f_{\rm esc} f_\ast$ than our fiducial choices in eq. \ref{eq:zeta} (see for example, Fig. 7 in \citealt{KF12}).
This is suggestive of at least a mild {\it halo mass} dependence of $f_{\rm esc} f_\ast$ (e.g. \citealt{AFT12}).}.  Further progress would be aided by improved, physically-motivated models for the evolution of the ionizing emissivity.

\vspace{0.5cm}

We thank M. McQuinn, M. Haehnelt, J. Bolton and S. Furlanetto for insightful comments on draft versions of this work.


\bibliographystyle{mn2e}
\bibliography{ms}

\end{document}